\documentclass[aps,prb,twocolumn,preprintnumbers,showpacs,amsmath,amssymb,floatfix]{revtex4}
\usepackage{latexsym}
\usepackage{dcolumn}
\usepackage[dvips]{graphicx}
\usepackage{amssymb}
\usepackage{graphics}
\usepackage{amsmath}
\usepackage{epsf}
\newcommand{\beq}{\begin{equation}}
\newcommand{\eeq}{\end{equation}}

\begin{document}
\title{Disorder-induced temperature-dependent transport in graphene: Puddles, impurities, activation, and diffusion}
\author{Qiuzi Li, E. H.\ Hwang, and S. Das Sarma}
\affiliation{Condensed Matter Theory Center, Department of Physics,
  University of Maryland, College Park, Maryland 20742}
\date{\today}

\begin{abstract}
We theoretically study the transport properties of both monolayer and bilayer
graphene in the presence of electron-hole puddles induced by charged
impurities which are invariably present in the graphene
environment. We calculate the graphene conductivity
by taking into account the non-mean-field two-component nature of transport in
the highly inhomogeneous density and potential
landscape, where activated transport across the potential fluctuations in the puddle regimes coexists with regular metallic diffusive transport. The existence of puddles allows the local activation
at low carrier densities,  giving rise to an insulating
temperature dependence in the conductivity of both monolayer and bilayer
graphene systems. We also critically study the qualitative similarity and the quantitative
difference between monolayer and bilayer graphene transport in the
presence of puddles. Our theoretical calculation explains the
non-monotonic feature of
the temperature dependent transport, which is experimentally generically observed in low mobility
graphene samples. We establish the 2-component nature (i.e., both activated and diffusive) of graphene transport arising from the existence of potential fluctuation induced inhomogeneous density puddles. The temperature dependence of the graphene conductivity arises from many competing mechanisms, even without considering any phonon effects, such as thermal excitation of carriers from the valence band to the conduction band, temperature dependent screening, thermal activation across the potential fluctuations associated with the electron-hole puddles induced by the random charged impurities in the environment, leading to very complex temperature dependence which depends both on the carrier density and the temperature range of interest.
\end{abstract}
\pacs{72.80.Vp, 72.10.-d, 73.22.Pr, 81.05.ue}
\maketitle

\section{Introduction}

Graphene, as a novel gapless two dimensional (2D) chiral electron-hole system, has
attracted great interest in recent years, both experimentally and
theoretically\cite{dassarma2010,castro2007}.
Its
transport properties have been at the center of key fundamental and
technological efforts with vast potential for applications in future
nanotechnology\cite{Youngbin_NanoLet10}. For monolayer graphene (MLG), the
fundamental interest arises from its unique linear chiral Dirac
carrier dispersion with a zero energy gap between conduction and valence
band\cite{Novoselov}. The bilayer graphene (BLG) is also intriguing as its physical
properties lie between
MLG and 2D semiconductor-based electron gas (2DEG) systems which
are gapped and non-chiral with a  quadratic band
dispersion.
Much of the early work on graphene transport
focused on the density-dependent (i.e., gate voltage
tuned)\cite{dassarma2010,Novoselov,TanDas_PRL07,Chen2008,ZhuExp_PRB09,Fang_Nano09} and
temperature-dependent
\cite{dassarma2010,YTan_EPJT07,Chen_NPHNano2008,HwangScreen_PRB09,Shaolong_PRB10} conductivity in
homogeneous MLG and BLG systems. The basic graphene transport properties, particularly at high densities far from the charge neutral Dirac point, are now reasonably well-understood\cite{dassarma2010}.

However, unintended charged impurities, which are invariably present
in the graphene environment, (e.g., the substrate-graphene interface), lead
to the formation of inhomogeneous electron-hole puddles in the system
\cite{EHwang_PRL07,RossiAdamDas_PRB09}, which have been confirmed by experiments\cite{Martin_NP08,YZhang_NP09}
using the techniques of
scanning potential and tunneling microscopies. Although MLG samples show a metallic behavior at high densities
a weak ``insulating" temperature-dependent
conductivity $\sigma(T)$ has been measured at low carrier density
and at the charge neutrality point
(CNP) \cite{YTan_EPJT07}. (We define
insulating/metallic temperature dependence of conductivity
$\sigma(T)$ as $d\sigma(T)/dT$ being positive/negative at fixed gate
voltage.) In addition, a recent experiment\cite{Heo_arXiv10} on low
mobility MLG grown by chemical vapor deposition (CVD) shows a
strong ``insulating" behavior at low temperatures and a metallic
feature at high temperatures manifesting a non-monotonic temperature
dependence in the measured electrical conductivity.
In BLG samples \cite{zhu2009,feldman2009,zhujun_PRBR10,Nam_PRB10} the strong
insulating behavior in the temperature dependent
conductivity has been observed not only near CNP but also at carrier densities as high as
$10^{12}$cm$^{-2}$ or higher. To be more specific, in
Ref.~[\onlinecite{zhu2009}], $\sigma(T)$ in BLG increases by $20-40\%$
as temperature $T$ increases from $4-300$ K for carrier density in the
range $3.19\times 10^{12}-7.16\times 10^{12}$ cm$^{-2}$. To understand
this anomalous temperature dependence in $\sigma(T)$, both MLG and
BLG, it is essential to know the role of disorder in graphene
transport. We note that phonon scattering (Ref.~[\onlinecite{HwangDasPhonon_PRB08,MinHwang_PRB11}]), although being weak in graphene, always contributes an increasing resistivity with increasing temperature and thus always leads to metallic behavior, and thus cannot be the mechanism for the intriguing insulating temperature dependence often observed in graphene transport at lower carrier densities -- in fact, at very high temperatures ($>300$ K) graphene should always manifest metallic temperature dependence in its conductivity due to phonon scattering effects which we would ignore in the current work.  Our goal here is to theoretically study in a comprehensive manner the temperature dependence of graphene transport properties arising entirely from the disorder effects.

The experimentally measured anomalous temperature
dependent conductivity of BLG has been theoretically investigated by
applying the analytic statistical theory to the inhomogeneous
potential fluctuation and it is found that the anomalous
BLG $\sigma(T)$ is likely to be caused by the electron-hole puddles
induced by randomly distributed disorder in the graphene
environment \cite{Hwang_InsuPRB2010}.
In this paper, we extend this work and apply the same analytic statistical
theory to MLG systems and explain the intriguing coexistence of both
metallic and insulating features of MLG $\sigma(T)$.
In the presence of
large fluctuating potentials $V({\bf r})$ associated with microscopic
configurations of Coulomb disorder in the system, the local Fermi level,
$\mu({\bf r}) = E_F - V({\bf r})$, would necessarily have large
spatial fluctuations. We carry out an analytical theory implementing
this physical idea by assuming that the value of the potential at any
given point follows a Gaussian distribution, parametrized by
$s=V_{rms}$ (the root-mean-square fluctuations or the standard
deviation in $V({\bf r})$ about the average potential). This
distribution can then
be used to average the local density of states to obtain effective
carrier densities, which can then be used to compute the physical
quantities of interest\cite{David_PRB10}.
The observed anomalous temperature dependent $\sigma(T)$ is then understood as the
competition between the thermal activation of carrier density and
temperature-dependent screening effects. Our theory explains the suppression of
the insulating behavior in higher mobility samples with lower
disorder, which is consistent with experimental observations.
We also provide the similarity and the quantitative
difference between monolayer and bilayer graphene transport in the
presence of puddles.

The motivation of our theory comes from the observation that the
electron-hole puddles, which dominate the low-density graphene
landscape, allow for a 2-component semiclassical transport behavior, where the
usual metallic diffusive carrier transport is accompanied by transport
by activated carriers which have been locally thermally excited above
the potential fluctuations imposed by the static disorder.  This
naturally allows for both insulating and metallic transport behavior
occurring preferentially respectively at lower and higher carrier
densities since the puddles disappear with increasing carrier density due to screening.
At zero temperature (where no activation is allowed) or at very high
carrier density (where puddles are suppressed), only diffusive
transport is possible.  But at any finite temperatures and at not too
high densities, there would always be a 2-component transport with
both activated and diffusive carriers contributing to conductivity.
Our theory develops this idea into a concrete description.  We
emphasize that our theory explicitly takes into account the
inhomogeneous nature of the graphene landscape and is non-mean-field
as a matter of principle.

This paper is organized as follows. In Sec.~\ref{sec:mden}, we
introduce the analytical statistical theory to describe random
electronic potential fluctuations created by charged impurities in the
environment. We also calculate the modified density of states and the
corresponding temperature-dependent effective  carrier density in
monolayer graphene. Then, in Sec.~\ref{sec:mono}, we describe the
calculations and the main features of the temperature-dependent
conductivity of MLG in the presence of density inhomogeneity. In
Sec.~\ref{sec:bden} and ~\ref{sec:bcon}, we elaborate and extend our
earlier results for the interplay between density inhomogeneity and
temperature in bilayer graphene (BLG) transport. We further discuss the connection
of our theory to earlier theories in Sec. \ref{sec:connection}. We discuss the similarities and quantitative differences among the effects of
inhomogeneity (i.e., the puddles) on MLG and BLG transport and
summarize our results in Sec.~\ref{sec:conclu}. In Appendix
\ref{sec:appen}, we discuss a microscopic theory to calculate the
effects of potential fluctuation on graphene systems, providing a
self-consistent formulation of graphene density of states in the
presence of random charged impurities near graphene/substrate
interface, showing in the process that this microscopically calculated density of states agrees well with the model density of states obtained from the Gaussian fluctuations.

\section{Temperature dependent carrier density for inhomogeneous MLG }
\label{sec:mden}

It is well known that MLG breaks up into an inhomogeneous landscape of
electron-hole puddles, especially around the charge neutral point
(CNP) \cite{Martin_NP08, Rossi_PRL08, YZhang_NP09}. Below we derive
an analytic statistical theory taking account of the effects of
inhomogeneous density in monolayer graphene (MLG) to explain the
nonmonotonic temperature dependent transport observed in MLG
\cite{YTan_EPJT07, Heo_arXiv10}.
We start by
assuming that charged impurities, located
in the substrate or near the graphene, create a local electrostatic
potential, which fluctuates randomly about its average value across
the surface of the graphene sheet. The potential fluctuations
themselves are assumed to be described by a statistical distribution
function $P(V)$ where $V=V({\bf r})$ is the fluctuating potential
energy at the point ${\bf r}\equiv (x,y)$ in the 2D MLG plane.
We approximate the probability $P(V)dV$ of finding the
local electronic potential
energy within a range $dV$ about $V$ to be a Gaussian form, i.e.,
\begin{equation}
P(V) = \frac{1}{\sqrt{2\pi s^2}} \exp(-V^2/2s^2),
\label{eq:fluc}
\end{equation}
where $s$ is the standard deviation (or equivalently, the strength of
the potential fluctuation), which is used as an adjustable parameter
to tune the tail-width\cite{Arnold_APL74}.  In the Appendix, we provide
a microscopic approach to self-consistently solve the strength of
potential fluctuations in the presence of charged impurities.
Due to the electron-hole symmetry in the problem, we only provide the formalism and equations for electron
like carriers and the hole part can be obtained simply by changing $E$ to
$-E$.

\begin{figure}
 \includegraphics[width=0.8\columnwidth]{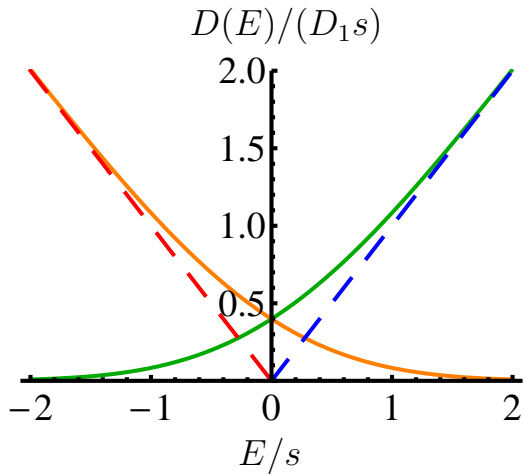}
\caption{ (Color online). Normalized density of states for both
  electron and hole in MLG. The solid and dashed lines are for the DOS
  in inhomogeneous and homogeneous systems, respectively. The
  electron (hole) band tail locates at $E<0$ ($E>0$), which gives rise
  to electron (hole) puddles at $E<0$ and $E>0$.
}
\label{fig:mdos}
\end{figure}

The  potential fluctuations
given by Eq.~(\ref{eq:fluc}) affect the overall electronic density of
states (DOS) in MLG. In our model we do not assume that the size of
  the puddles to be identical, but we take the puddle sizes to be completely
  random controlled by the distribution function given in Eq. \ref{eq:fluc}.  We emphasize that our assumption of a Gaussian distribution for the potential fluctuations, equivalently implying a Gaussian distribution for the density fluctuations associated with the puddles, is known to be an excellent quantitative approximation to the actual numerically calculated puddle structures in graphene\cite{dassarma2010,Rossi_PRL08}. The characteristics of the puddles are determined by both the sign
  and the magnitude of  $V-E_F$, i.e., a negative (positive) $V-E_F$
  indicates an electron (hole) region.  A different approach utilizing equal
  size puddles with a certain potential $V$ has been used to
  calculate transport coefficients using a numerical transfer matrix technique
  \cite{Jose_PRB07}.
Then in the presence of electron-hole puddles
the density of states  is increased by the allowed electron region
fraction and given by \cite{zallen1971,eggarter1970,Arnold_APL74}
\begin{eqnarray}
\begin{array}{l l l l l l l l }
D_e(E) = \int _{-\infty }^E\dfrac{g_s g_v (E-V)}{2\pi (\hbar v_F)^2}P(V) dV
\\
=D_1 \big[\dfrac{E}{2} \text{erfc}(-\dfrac{E}{\sqrt{2}
    s})+\dfrac{s}{\sqrt{2 \pi}} \exp(-\dfrac{E^2}{2s^2})\big],
\label{eq:mdos}
\end{array}
\end{eqnarray}
where erfc$(x)$ is the
complementary error function,
\begin{equation}
\text{erfc}(x) = \dfrac{2}{\sqrt{\pi }}\int_{x}^{\infty}e^{-t^2}dt,
\label{eq:erfc}
\end{equation}
and $D_1 = \dfrac{g_s g_v}{2\pi (\hbar v_F)^2}$,
where $v_F$ is the graphene Fermi velocity, $g_s=2$ and $g_v=2$ are the spin and
valley degeneracies, respectively.
We have $D_1=1.5\times 10^{8}$ cm$^{-2}$/meV$^{2}$ with the Fermi velocity
$v_F=10^{6}$ m/s.
Note that the tail of the DOS is determined by the potential fluctuation
strength $s$. For the case $s=0$, the system becomes homogeneous and $D_e(E) = D_1 E$.
In this case there is no carrier density at Dirac point ($E=0$) at zero
temperature. It is apparent that in the presence of potential
fluctuations, the $D_e(E)$ starts at finite value $\frac{D_1 s}{\sqrt{2
    \pi}}$ at $E=0$ and approaches $D_1 E$ in high energy limit. For
high-energy limit, the carrier is essentially free
since nearly every point of the system is accessible.  In
Fig. \ref{fig:mdos}, we show the normalized density of states as a
function of energy for both electrons and holes in MLG.
We mention that the self-consistent microscopic theory gives the same structure for
the density of states of graphene systems (see
Appendix~\ref{sec:appen}).

Since monolayer graphene is a semi-metal or zero-gap semiconductor,
the electron density at finite temperatures increases due to the direct
thermal excitation from valence band to conduction band, which is one
of the
important sources of temperature dependent transport at low carrier
densities. Therefore, we
first consider the temperature dependence of thermally
excited electron density.
The total electron density is given by
\begin{equation}
n_e  = \int_{-\infty}^{\infty}D_e(E)\frac{dE}{e^{\beta(E-\mu)}+1},
\label{eq:mdenint}
\end{equation}
where $\beta=1/k_BT$ and $\mu$ is the chemical potential. At $T=0$, $\mu$ becomes the Fermi energy $\mu(T=0)=E_F$.

\subsection{$n_e (T)$ of MLG at CNP ($E_F = 0$)}

\begin{figure}
\epsfysize=1.8in
\epsffile{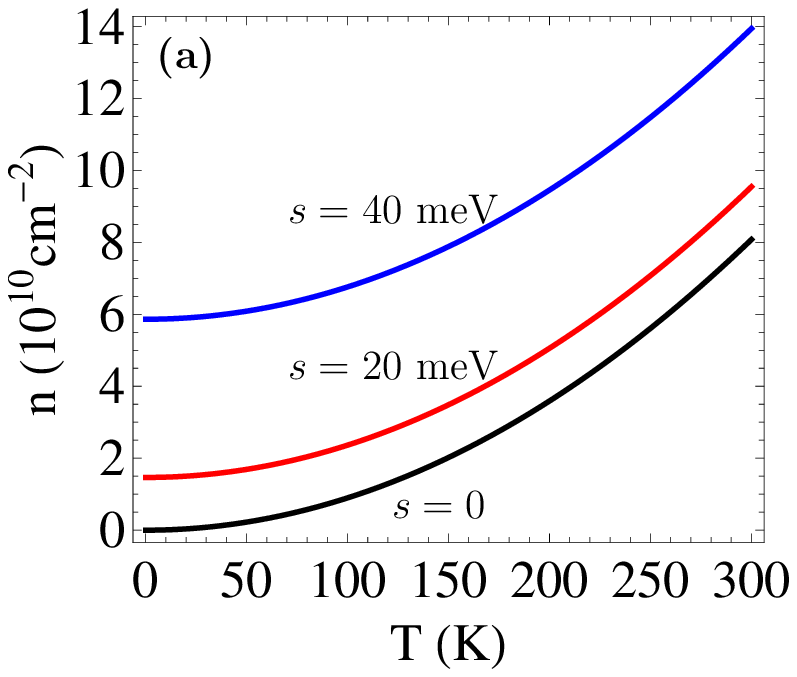}
\epsfysize=1.8in
\epsffile{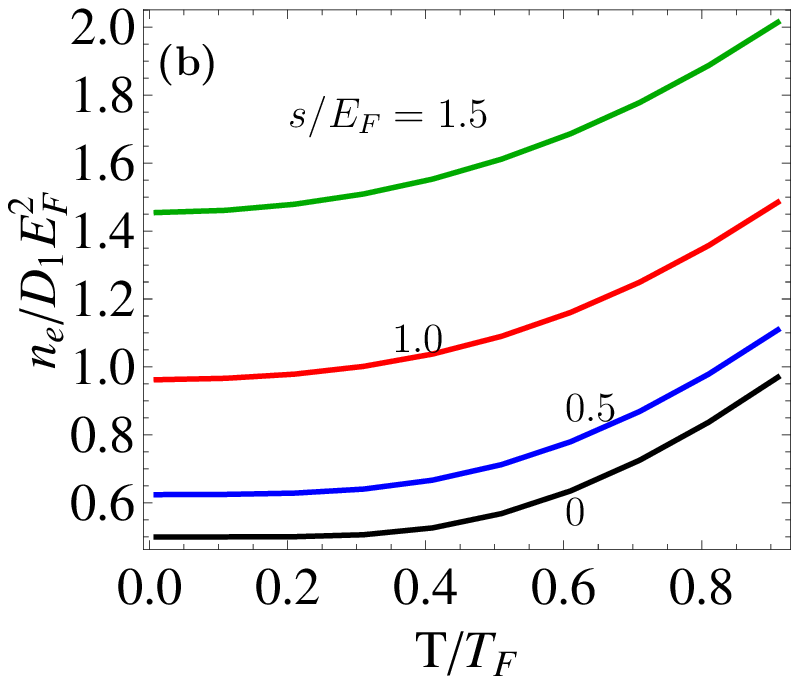}
\epsfysize=1.8in
\epsffile{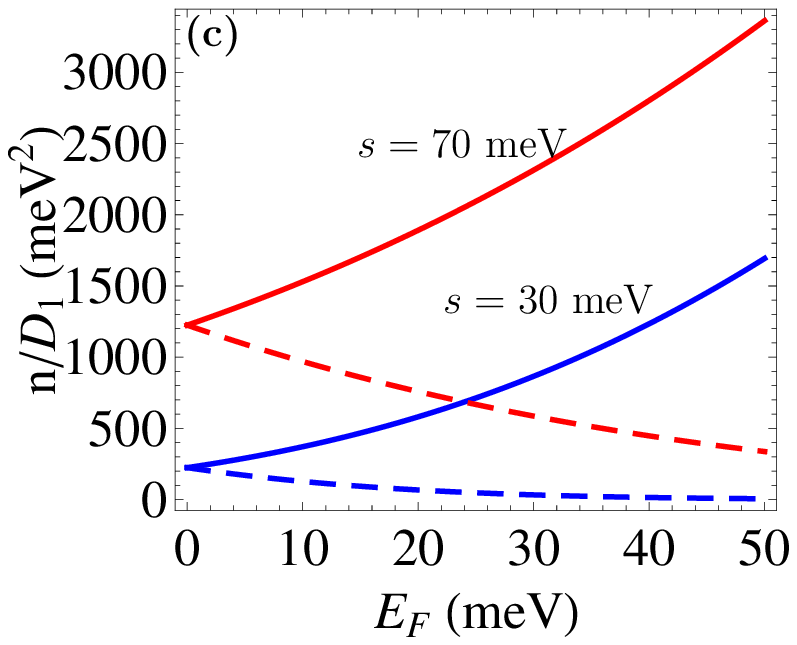}
\caption{ (Color online)
(a) The electron density of MLG at CNP as a function of
  temperature for different $s$. At $T=0$ the density is given by
  $n_0=D_1 s^2/4$.
(b) The temperature dependent electron density of MLG at finite $E_F$
for different $s$. For $s/E_F \neq 0 $
  the leading order behavior is quadratic.
(c) Total electron densities (solid lines) and hole densities (dashed
lines) of MLG as a function of $E_F$ for two
different $s=30$ meV and 70 meV. The densities at the band tails are given by $n_e(E_F=0)=n_h(E_F=0)=D_1
s^2/4$.
}
\label{fig:den}
\end{figure}

When the Fermi energy is zero (or at CNP) all
electrons are located in the band tail at $T=0$ and
the electron and hole densities in the band tail are given by
\begin{equation}
n_0=n_e(E_F=0)=n_h(E_F=0)= D_1 \frac{s^2}{4}.
\end{equation}
Note that the electron (or hole) density in the band tails increases
quadratically with the standard deviation $s$.
At finite temperatures the behavior of $n_e(T)$ at
CNP becomes
\begin{equation}
n_e(T) = n_0 \left [ 1 + \frac{\pi^2}{3} \left ( \frac{k_BT}{s} \right
  )^2 \right ].
\label{eq:den_0}
\end{equation}
The leading order temperature dependence in $n_e(T)$ is quadratic.
For homogeneous MLG ($s=0$) with the linear-in-energy behavior of the
DOS, the electron density is given by
$n_e(T)=\dfrac{D_1 \pi^2}{12} k_B^2 T^2$.
 In particular, in the ballistic regime the number of propagating
  channels increases due
  to the thermal smearing of the Fermi surface, which leads to
  the observation of an insulating behavior in $\sigma(T)$ at CNP for
  high mobility suspended graphene
  samples\cite{Du_NaNano08,Bolotin_RPL08,Muller_PRL09}.
The presence of the band tail does not change the quadratic
temperature dependence in the thermal excitation
when the system is at the charge neutral point ($E_F =
0$). But the inhomogeneous MLG has $n_0$ electrons in the band tails.
In Fig.~\ref{fig:den}(a) we
show the temperature dependent electron
density at CNP for different values of standard deviation $s$.

\subsection{$n_e (T)$ of MLG at finite doping ($E_F > 0$)}

In the case of finite doping (or gate voltage), i.e., $E_F\neq 0$,
the
electron density of the homogeneous MLG (i.e., $s=0$) is given by
\begin{eqnarray}
\begin{array}{l l l l l l l l }
n_{0 e} (T) = D_1 \int _{0}^\infty \dfrac{E dE}{\exp(\beta (E-\mu_0))+1}
\\
\ \ \ \ \ \ \ \ \ \ =-D_1 \dfrac{F_1 (\mu_0 \beta)}{\beta^2}
\label{Eq:hf}
\end{array}
\end{eqnarray}
where $F_1 (x) = \int _{0}^{\infty} \dfrac{t \ dt}{1+ \exp(t-x)}$, and
$\mu_0$ is the chemical potential of homogeneous MLG and is
determined by the conservation of the total electron density.
Then the chemical potential is given by the following relation,
$\dfrac{E_F^2 \beta^2 }{2}=F_1 (\beta \mu_0)-F_1 (-\beta \mu_0)$.
Using the asymptotic forms \cite{HwangScreen_PRB09} of the function
$F_1(x)$ for $x \ll 1$ and $x \gg 1$, i.e.,
\begin{equation}
\begin{array}{c c }
F_1 (x) \approx \dfrac{\pi^2}{12}+x \text{ln}2+\dfrac{x^2}{4} \ \ \ \text{for}\ \ |x|\ll 1
\\
\\
F_1 (x) \approx \left[\dfrac{x^2}{2}+\dfrac{\pi^2}{6}\right]\theta(x)+x \text{ln}(1+e^{-|x|})\ \text{for}\ |x|\gg 1,
\end{array}
\end{equation}
we have the asymptotic formula for the chemical potential in both low-
and high-temperature limits for homogeneous MLG
\begin{equation}
\begin{array}{c c }
\mu_0 (T) \simeq E_F \Big[1- \dfrac{\pi^2}{6}(\dfrac{T}{T_F})^2\Big] \ \ \ \text{for}\ \ T \ll T_F
\\
\\
\mu_0 (T) \simeq \dfrac{E_F}{4 \ln2}\dfrac{T_F}{T} \ \ \ \text{for}\ \ T \gg T_F.
\label{eq:mhomochem}
\end{array}
\end{equation}
Then the corresponding asymptotic formula of
the electron density (Eq. \ref{Eq:hf}) are given by
\begin{equation}
\begin{array}{c c }
n_{0e}(T) \simeq  \dfrac{D_1 E_F^2}{2}\left(1+ \dfrac{\pi^4}{36}\dfrac{T^4}{T_F^4}\right) \ \ \ \text{for}\ \ T \ll T_F
\\
\\
n_{0e}(T) \simeq  \dfrac{D_1 E_F^2 \pi ^2}{12 }\dfrac{T^2}{T_F^2} \ \ \text{for}\ \ T \gg T_F
\label{eq:mhomden}
\end{array}
\end{equation}
Since the direct thermal excitation is suppressed due to the finite Fermi energy, the excited electron density at low temperatures ($T \ll T_F$)
increases quartically rather than quadratically. But at high
temperatures ($T \gg T_F$), the total electron density
becomes a quadratic function of temperature as shown for an undoped MLG.

Next, we derive the temperature dependence of thermally excited
electron density in
the presence of electron-hole puddles ($s \neq 0$) at finite doping
($E_F\neq 0$). At zero temperature the
electron density for the inhomogeneous MLG can be written as:
\begin{eqnarray}
\begin{array}{l l l l l l l l }
n_e (0) =\dfrac{D_1 E_F^2}{4} \big[(1+ \tilde{s}^2)\text{erfc} (-\dfrac{1}{\sqrt{2}\tilde{s}})+ \sqrt{\dfrac{2}{\pi}} \tilde{s} \exp(-\dfrac{1}{2\tilde{s}^2})\big]
\\
\\
n_h (0) =\dfrac{D_1 E_F^2}{4} \big[(1+ \tilde{s}^2)\text{erfc} (\dfrac{1}{\sqrt{2}\tilde{s}})- \sqrt{\dfrac{2}{\pi}} \tilde{s} \exp(-\dfrac{1}{2\tilde{s}^2})\big]
\end{array}
\end{eqnarray}
where $\tilde{s}=s/E_F$.
The presence of electron-hole
puddles does not induce any additional charge in the
MLG system and the net carrier density $n=n_e-n_h$ should be
conserved. Then, the finite temperature chemical potential $\mu(T)$
changes as a function of both temperature and the strength of
potential fluctuation $s$, and it should satisfy the following
relation:
\begin{equation}
\begin{array}{l l l l l l l l }
\int _{-\infty}^{E_F} D_e(E) dE \ - \int _{E_F}^{\infty} D_h(E) dE
\\
\\
= \int _{-\infty}^{\infty} \dfrac{D_e(E) dE}{1+\exp(\beta (E-\mu))} \ - \int _{-\infty}^{\infty} \dfrac{D_h(E) dE}{1+\exp(\beta (\mu-E))},
\end{array}
\end{equation}
where $D_e(E)$ is the electronic density of states given by
Eq.~\ref{eq:mdos} and $D_h (E) = D_e(-E)$ is the density of states for
holes. The asymptotic analytical formula of  the chemical potential
$\mu (T)$ for inhomogeneous MLG is obtained as:
\begin{equation}
\begin{array}{c c }
\mu (T) \simeq E_F \Big[1- \dfrac{\pi^2}{6}(\dfrac{T}{T_F})^2 A(\tilde{s})\Big] \ \ \ \text{for}\ \ T \ll T_F
\\
\\
\mu (T) \simeq E_F B (\tilde{s},t)\ \  \ \text{for}\ \ \ \  T \gg T_F
\label{eq:mchem}
\end{array}
\end{equation}
 where functions $A(\tilde{s})$ and $B(\tilde{s})$ are given as follows:
\begin{equation}
\begin{array}{l l }
A(\tilde{s}) = e^{\frac{1}{2 \tilde{s}^2}}
\text{erf}\big[\dfrac{1}{\sqrt{2}
    \tilde{s}}\big]\Big/\big(\sqrt{\dfrac{2}{\pi }}
\tilde{s}+e^{\frac{1}{2 \tilde{s}^2}}
\text{erf}\big[\dfrac{1}{\sqrt{2} \tilde{s}}\big]\big)
\\
\\
B (\tilde{s}, t)=\Big(\dfrac{e^{-\frac{1}{2 \tilde{s}^2}} \tilde{s} }{ \sqrt{2 \pi }}+\dfrac{1}{2} (\tilde{s}^2+1) \text{erf}\big[\dfrac{1}{\sqrt{2} \tilde{s}}\big]\Big)\Big/\big(2 \ln2 \ t +\dfrac{\tilde{s}^2}{4 t}\big)
\end{array}
\end{equation}
where  $t= T/T_F$ and $\text{erf}(x) = \frac{2}{\sqrt{\pi
}}\int_{0}^{x}e^{-t^2}dt$ is the error function.

Combining Eqs. \ref{eq:mdos}, \ref{eq:mdenint} and \ref{eq:mchem}, we
obtain the asymptotic analytical formula of the electron density
for inhomogeneous MLG at low- and high-temperature limits as:
\begin{eqnarray}
\begin{array}{l l l l l l l l }
n_e (T)\simeq n_e (0)+ D_1 E_F^2 \dfrac{\pi ^2 }{12 }  \dfrac{T^2}{T_F^2}(1-A(\tilde{s})) \ \ \ \text{for}\ \ T \ll T_F
\\
\\
n_e (T)\simeq n_{0e}(T)+ n_e (0)-\dfrac{D_1 E_F^2}{4} \  \ \text{for}\ \ \ \  T \gg T_F
\label{eq:den_mu}
\end{array}
\end{eqnarray}
In the low temperature limit ($T \ll T_F$), the leading order term for the electron density has the same quadratic behavior as in undoped homogeneous MLG ($E_F=0$), but the coefficient is strongly
suppressed by fluctuation for the case of $s < E_F$, i.e., the high carrier density sample.  While in the case of $s > E_F$, i.e., the low carrier density sample,
the existence of electron-hole puddles gives rise to a notable
quadratic behavior for electron density $n_{e} (T)$ [see Fig.~\ref{fig:den}(b)].

\section{Conductivity of inhomogeneous MLG}
\label{sec:mono}

In this section,  we calculate the finite temperature conductivity for
inhomogeneous MLG with the temperature-dependent effective carrier
density derived above. The existence of electron-hole puddles allows
that the current flows through ``percolation channels" and the
transport properties of the inhomogeneous MLG system can be derived
using  the self-consistent effective medium theory of conductance in composite
mixtures\cite{kirkpatrick1973}, where the number of electrons per puddle
is not an important issue for our theory. The
percolation assumption is valid as long as the potential
fluctuation is larger than the thermal energy of the
carriers. Otherwise transport due to disorder scattering dominates. We emphasize that in our formalism the crossover from the percolation transport to ordinary scattering-dominated diffusive transport is guaranteed as the temperature is increased since we are explicitly taking into account both diffusive transport of free carriers and activated transport of the classically-localized carriers in our theory.  The only effects we neglect are quantum tunneling through the potential barriers and quantum interference since ours is a semiclassical theory. We also do not consider Klein tunneling explicitly in this paper because the Klein tunneling occurs at zero temperature for normal incident carriers at the electron-hole puddle boundary.
We also apply the Boltzmann transport
theory, where we  include the scattering mechanism with screened
Coulomb impurities and short-range disorder\cite{HwangScreen_PRB09}.
Note that
 the application of Boltzmann transport theory is justifiable because
 the quantum interference effects are not experimentally observed in
 the temperature regime of interest to us in this work.  It is conceivable that quantum interference and localization play some roles in graphene transport at very low temperatures, which is beyond the scope of this paper.  We also neglect all phonon effects in this work since electron-phonon coupling is weak in graphene.  Phonon effects are relevant at high temperatures ($>100$ K) and have been considered in the literature\cite{HwangDasPhonon_PRB08,MinHwang_PRB11}.

At CNP ($E_F=0$) electrons and holes are
equally occupied. As the Fermi energy increases, more electrons occupy
increasingly larger proportion of space. As the Fermi energy increases
to $E_F \gg s$, nearly all space is populated by the electrons [see
Fig.~\ref{fig:den}(c)] and the conductivity of the system approaches
the characteristic of the homogeneous material. Thus, there is a
possible coexistence of metallic and thermally-activated transport in
the presence of electron-hole puddles. When electron puddles occupy
more space than hole puddles, most electrons follow the continuous
metallic paths
extended throughout the system, but it is possible at finite
temperatures that the thermally activated transport
of electrons persists above the hole puddles. On the other
hand, holes in hole puddles propagate freely, but when
they meet electron puddles, activated holes conduct over
the electron puddles. Carrier transport in each puddle
is characterized by propagation of weak scattering transport theory\cite{kirkpatrick1973}. The activated carrier transport of prohibited regions, where the local potential energy $V$ is
less (greater) than Fermi energy for electrons (holes), is
proportional to the Fermi factor. If $\sigma_e$ and $\sigma_h$ are
the average conductivity of electron and hole puddles,
respectively, then
the activated conductivities are given by
\begin{subequations}
\begin{eqnarray}
\sigma_e^{(a)}(V) & = & \sigma_e \exp[\beta (E_F-V)], \\
\sigma_h^{(a)}(V) & = &\sigma_h \exp[\beta (V-E_F)],
\end{eqnarray}
\end{subequations}
where the density and temperature dependent average conductivities
($\sigma_e$ and $\sigma_h$) are
given within the Boltzmann transport theory \cite{dassarma2010} by
$\sigma_{e} \propto n_e \langle \tau \rangle$ and
$\sigma_{h} \propto n_h \langle \tau \rangle$,
where $n_e$ and $n_h$ are average electron and hole densities,
respectively, and $\langle \tau \rangle$ is the average transport relaxation time which includes the thermal smearing effects and depends explicitly on the scattering mechanism
\cite{dassarma2010}
and it is given by,
\begin{equation}
\langle \tau \rangle = E_F\dfrac{\int d\epsilon D_e(\epsilon) \tau(\epsilon)(-\partial f/\partial\epsilon)}{\int d\epsilon D_e(\epsilon) f(\epsilon)}
\label{eq:avtau}
\end{equation}
where $\tau(\epsilon)$ and $f=1/(1+e^{\beta(\epsilon-\mu)})$ are,
respectively, the energy-dependent transport scattering time and the
finite temperature Fermi distribution function. Because the density
inhomogeneity effects already been considered in the variation of
effective carrier density, we use the DOS
of homogeneous MLG $D_e(\epsilon) = D_1 \epsilon$ in
Eq.~\ref{eq:avtau} to avoid double counting. And $\tau(\epsilon)$ is
given by
\begin{eqnarray}
\dfrac{\hbar}{\tau(\epsilon_{p{\bf k}})}= 2\pi n_{dis} \int \frac{d^2 k'}{(2\pi)^2}|\langle V_{p{\bf k},p{\bf k'}}\rangle|^2 g(\theta_{\bf kk'})\nonumber \\
\times \left[1-\cos\theta_{\bf kk'}\right]\delta(\epsilon_{p\mathbf{k'}}-\epsilon_{p\mathbf{k}})
\label{eq:mscatt}
\end{eqnarray}
where $\epsilon_{p\mathbf{k}} = p \hbar v_F k$ is the carrier energy
for the pseudospin state ``$p$" and ${\bf k}$ is the 2D wave vector,
$\langle V_{p{\bf k},p{\bf k'}}\rangle$ is the matrix element of the
impurity disorder potential in the system environment, $\theta_{\bf kk'}$ is
the scattering angle between in- and out- wave vectors
${\bf k}$ and $\bf k'$, $g(\theta_{\bf kk'})=\left[1+\cos\theta_{\bf
    kk'}\right]/2$ is a wave function form factor associated with the
chiral nature of MLG (and is determined by its band
structure). $n_{dis}$ is the appropriate 2D areal concentration of the
impurity centers giving rise to the random disorder
potential\cite{DasEnrico_PRB10}. We consider two different kinds of
disorder scattering mechanisms: (i) randomly distributed screened
Coulomb disorder for which $n_{dis} |\langle V_{p{\bf k},p{\bf
    k'}}\rangle|^2 = n_i |v_i(q)/\varepsilon(q)|^2$, where $v_i (q) =
2 \pi e^2/(\kappa q)$ is the Fourier transform of the 2D Coulomb
potential in an effective background lattice dielectric constant
$\kappa$ and $\varepsilon(q)\equiv \varepsilon(q,T)$ is the 2D finite
temperature static RPA dielectric function\cite{HwangDas_PRB07} (Note that we use $n_i$ to
denote the charged impurity density); (ii) short-range disorder for
which $n_{dis} |\langle V_{p{\bf k},p{\bf k'}}\rangle|^2 = n_d V^2_0$
where $n_d$ is the 2D impurity density and $V_0$ is a constant short-range
(i.e. a $\delta$-function in real space) potential strength.
Note that the use of Born approximation for short-range disorder
  requires weak scattering condition\cite{Ferreira_PRB11}, which is
  verified by the disorder parameters we use in our calculation.

Now we denote the electron (hole) puddle as region `1' (`2').
In region 1 electrons are occupied more space than holes when $E_F>0$.
The fraction of the total area occupied by electrons with
Fermi energy $E_F$ is given by
$p=\int_{-\infty}^{E_F}P(V)dV$.
Then the
total conductivity of region 1 can be calculated,
\begin{eqnarray}
\sigma_1 & = & \frac{1}{p}\int^{E_F}_{-\infty}(\sigma_e +
\sigma_h^{(a)})P(V) dV, \nonumber \\
&=& \sigma_{e}+\frac{\sigma_{h}}{2p} e^{
    \frac{\beta^2s^2}{2} -\beta E_F } {\rm erfc} \left (
    -\frac{E_F}{\sqrt{2} s} + \frac{\beta s}{\sqrt{2}} \right).
\label{eq:sig1}
\end{eqnarray}
At the same time the holes occupy the area with a fraction $q=1-p$ and
the total conductivity of region 2 becomes
\begin{eqnarray}
\sigma_2 & = &\frac{1}{q}\int_{E_F}^{\infty}(\sigma_e^{(a)} + \sigma_h)P(V) dV
\nonumber \\
&=& \sigma_{h}+\frac{\sigma_{e}}{2q} e^{
    \frac{\beta^2s^2}{2} +\beta E_F } {\rm erfc} \left (
    \frac{E_F}{\sqrt{2} s} + \frac{\beta s}{\sqrt{2}} \right).
\label{eq:sig2}
\end{eqnarray}
The $\sigma_1$ and $\sigma_2$ are distributed according to the
binary distribution.
The conductivity of binary system can be calculated by using
the effective medium theory of conductance in
mixtures\cite{kirkpatrick1973}. The result
for a 2D binary mixture of components with conductivity $\sigma_1$ and
$\sigma_2$ is given by \cite{kirkpatrick1973}
\begin{equation}
\sigma_t = (p-\frac{1}{2})\left [ (\sigma_1 -\sigma_2) +
  \sqrt{(\sigma_1-\sigma_2)^2+\frac{4\sigma_1 \sigma_2}{(2p-1)^2}}
  \right ].
\label{eq:sig_tot}
\end{equation}
This result can be applied for all Fermi energy. For a large doping
case, in which the hole puddles disappear, we have $p=1$ and
$\sigma_2=0$, then Eq.~(\ref{eq:sig_tot}) becomes $\sigma = \sigma_1$,
i.e., the conductivity of electrons in the homogeneous system.

\begin{figure}
\includegraphics[width=6.5cm]{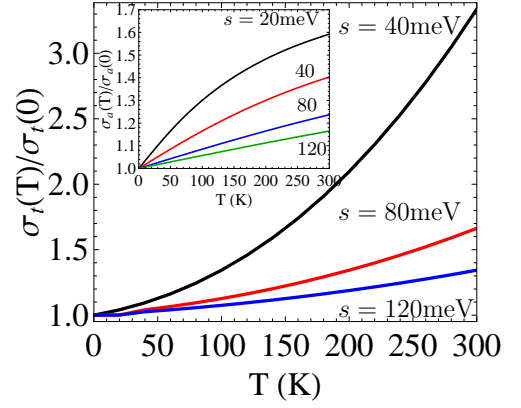}
\caption{(Color online)
$\sigma_t(T)$ of MLG at charge neutral point for
different $s$ (Eq.~\ref{eq:msig1} and $n_e$ as given in Eq.~\ref{eq:den_0}). Inset shows the thermally activated conductivity of MLG as a
function of temperature, where $\sigma_a(T)/\sigma_a(0)=1+ e^{\beta^2 s^2/2}
  {\rm erfc} (\beta s/\sqrt{2}) $.
\label{fig:sig_mu0}
}
\end{figure}

\subsection{$\sigma(T)$ of MLG at CNP ($E_F = 0$)}

We first consider the conductivity at CNP ($E_F=0$). The
conductivities in each region are given by
\begin{subequations}
\begin{eqnarray}
\sigma_1 & = &  \sigma_{e} \left [ 1 + \frac{\eta}{2p} e^{\beta^2 s^2/2}
  {\rm erfc} (\beta s/\sqrt{2}) \right ],  \label{eq:msig1} \\
\sigma_2  & = & \sigma_{h} \left [ 1 + \frac{1}{2q\eta} e^{\beta^2 s^2/2}
  {\rm erfc} (\beta s/\sqrt{2}) \right ],
\end{eqnarray}
\end{subequations}
where $\eta = n_h/n_e$ is the ratio of the hole density to the
electron density.
Since the electrons and holes are equally populated, we have $p=q=1/2$
and $\sigma_{e}=\sigma_{h}$, then the total conductivity becomes
$\sigma_{t}  = \sqrt{\sigma_1 \sigma_2} = \sigma_1$.
The asymptotic behavior of the conductivity at low temperatures ($k_BT
\ll s$) becomes
\begin{equation}
\sigma_t(T) = \sigma_{e} \left [1 + \sqrt{ \frac{2}{\pi}} \frac{k_BT}{s}
  - \frac{2}{\sqrt{\pi}}\frac{(k_BT)^3}{s^3} \right ].
  \label{eq:siglast}
\end{equation}
The activated conductivity increases linearly with a slope
$\sqrt{2/\pi}k_B/s$ as temperature increases. Typically
$s$ is smaller in higher mobility samples, which gives rise to stronger
insulating behavior at low
temperatures. The next order temperature correction to  conductivity
arises from the thermal excitation given in Eq.~(\ref{eq:den_0}) which
gives quadratic ($T^2$) temperature corrections. Thus, in the low
temperature limit the total conductivity at the
CNP is given by:
\begin{equation}
\sigma_t = \sigma(0) \left [1+ \sqrt{\frac{2}{\pi}}\frac{k_BT}{s} +
  \frac{\pi^2}{3} \left ( \frac{k_BT}{s} \right )^2 \right ].
\end{equation}
At high temperatures ($k_BT \gg s$) we have
\begin{equation}
\sigma_t = \sigma_e \left [ 2 - \sqrt{ \frac{2}{\pi}} \frac{s} {k_BT} +
  \frac{s^2}{2 (k_BT)^2} \right ].
\end{equation}
where the temperature dependence of $\sigma_e$ has been given in Eq.~(\ref{eq:den_0}). The total conductivity due to the activation behavior approaches
a limiting value and all temperature dependence comes from the thermal
excitation through the change of the effective carrier density in the presence
of the inhomogeneity given in
Eq.~(\ref{eq:den_0}). Thus at very high temperatures ($T\gg s/k_B$)
the MLG conductivity at the charge neutral point increases quadratically regardless of
the sample quality. In Fig.~\ref{fig:sig_mu0} the
temperature dependent conductivity  has been calculated at charge
neutral point, where the temperature dependent scattering mechanism
can be neglected. In Ref. [\onlinecite{Heo_arXiv10}],
about 60\% increase of conductivity is observed as the temperature
increases from 4 K to 300 K.  We estimate the potential fluctuation parameter
$s\sim 80 $ meV for this sample based on our theoretical analysis as compared with the data.

\subsection{$\sigma(T)$ of MLG at finite doping ($E_F > 0$)}

At finite doping ($E_F > 0$) the temperature dependent conductivities
are very complex because three energies ($E_F$, $s$, and
$k_BT$) are competing among them. Especially when $k_BT \ll s$, regardless of
$E_F$, we have the asymptotic behavior of
conductivities in region 1
and 2 from Eqs.~(\ref{eq:sig1}) and (\ref{eq:sig2}), respectively,
\begin{subequations}
\begin{eqnarray}
\sigma_1 & = & \sigma_{e} \left [ 1+ \frac{\eta}{2p} e^{-1/2\tilde{s}^2}
\sqrt{\frac{2}{\pi}}  \frac{1}{\tilde{s}/t-1/\tilde{s}} \right ], \\
\sigma_2 & = & \sigma_{h} \left [ 1+ \frac{1}{2q \eta} e^{-1/2\tilde{s}^2}
\sqrt{\frac{2}{\pi}}  \frac{1}{\tilde{s}/t+1/\tilde{s}} \right ],
\end{eqnarray}
\label{eq:msigt}
\end{subequations}
where $\tilde{s}=s/E_F$ and $t=T/T_F$. The leading order correction is
linear but the coefficient is exponentially suppressed by the term
$\exp(-E_F^2/2s^2)$. This fact indicates that in the high mobility
sample with small $s$, the activated conductivity is weakly
temperature dependent except at low density regimes, i.e. $E_F < s$.
Since the density increase by thermal excitation is also suppressed
exponentially by the same factor [see Eq.~(\ref{eq:den_mu})], the
dominant temperature dependent conductivity arises from the scattering time \cite{dassarma2010}, which manifests the metallic behavior. On the other hand,
for a low mobility sample with a large $s$, the linear temperature
dependence due to thermal activation can be observed even at high carrier
densities $E_F \agt s$.
\begin{figure}
\epsfysize=2.0in
\epsffile{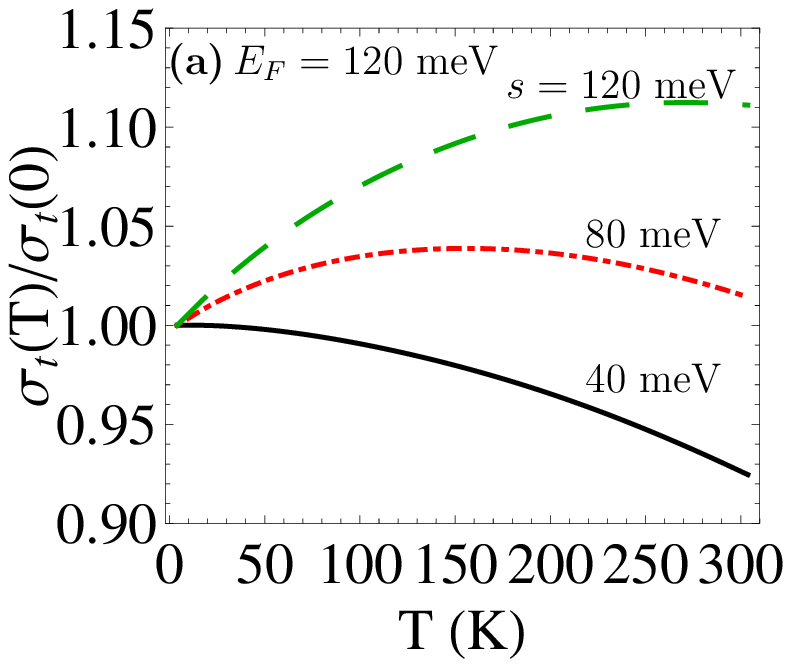}
\epsfysize=2.0in
\epsffile{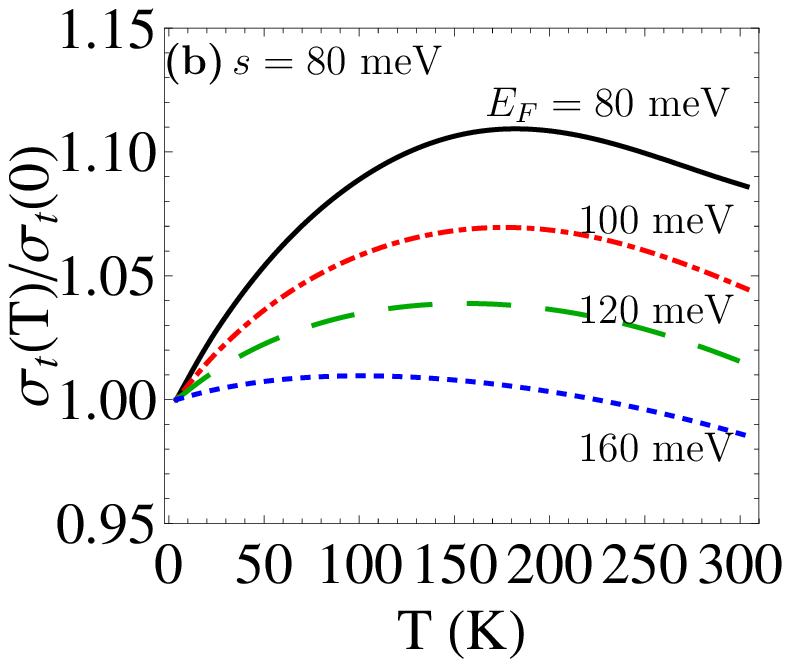}
  \caption{(Color online). Calculated total conductivity $\sigma_t(T)/\sigma_t(0)$ of MLG with the following parameters: $n_i = 10^{12}$ cm$^{-2}$ and $n_d V_0^2 = 2$ (eV {\AA})$^2$. (a) $\sigma_t(T)$ for  $E_F=120$ meV and for
    different $s$. (b) $\sigma_t(T)$ of MLG
for $s=80$ meV and for several $E_F=80$, 100, 120, 160 meV, which correspond to the net carrier densities $n = n_e-n_h \simeq 0.9\times 10^{12}$, $1.2\times 10^{12}$, $1.5\times 10^{12}$, and $2.3\times 10^{12}$ cm$^{-2}$.}
  \label{fig:sig_act}
\end{figure}

In Fig.~\ref{fig:sig_act} we present the total conductivities of
inhomogeneous MLG as a function of temperature (a) for a fixed Fermi
energy and several $s$ and (b)
for a fixed $s$ and several Fermi energies. The calculations for
Fig. \ref{fig:sig_act} are all carried out for MLG on SiO$_2$
substrate (corresponding to dielectric constant $\kappa \approx 2.5$),
charged impurity density $n_{i} = 10^{12}$ cm$^{-2}$ and short-ranged
disorder strength $n_d V_0^2 = 2$ (eV {\AA})$^2$. For total
conductivity, the thermally activated insulting behavior competes with
the temperature-dependent screening effects, where the latter always
give the metallic behavior in conductivity for MLG samples. When $s$
is small, the activated behavior is suppressed and the total
conductivity shows the metallic behavior. While for large value of
$s$, i.e., the low mobility sample, the thermal activation overwhelms
the metallic temperature dependence and the system manifests insulating
behavior. For $s \sim E_F$ the situation becomes much complex. At low temperatures,
the leading order of the temperature dependence is linear (the second term
in Eq. (\ref{eq:msigt})) and the total conductivity starts at weakly
insulating behavior. As the temperature increases, the screening
effects begin to dominant leading to the metallic behavior. As a result,
the temperature evolution of the conductivity becomes non-monotonic and
for large $s$ (or low mobility samples) the nonmonotonic behavior can
be more pronounced as shown in experiments \cite{Heo_arXiv10}.

\section{Temperature dependent carrier density of inhomogeneous BLG }
\label{sec:bden}

In the following of this paper, we extend our previous
study\cite{Hwang_InsuPRB2010} on the insulating behavior in metallic
bilayer graphene and compare it with MLG situation. The most
important difference between MLG and BLG comes from the fact  that, in
the BLG,  the two layers are weakly coupled by interlayer tunneling,
leading to an approximately parabolic band dispersion with an
effective mass about $m \simeq 0.033m_e$ ($m_e$ corresponds to the
bare electron mass) contrast to linear-dispersion Dirac carrier system
for MLG.  As done for MLG, we assume the electronic potential
fluctuations in BLG system to be a Gaussian form given in
Eq. \ref{eq:fluc} and this potential is felt equally by both
layers\cite{David_PRB10}.

\begin{figure}
 \includegraphics[width=0.8\columnwidth]{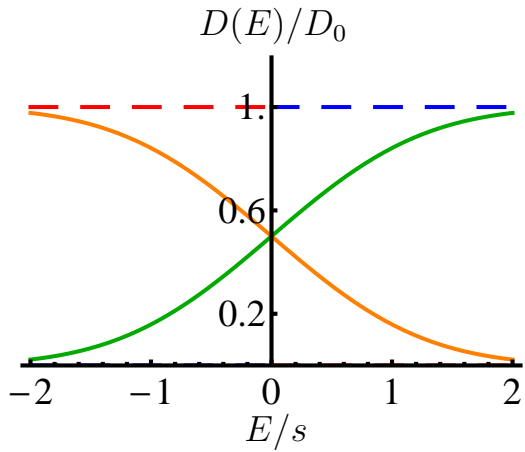}
\caption{ (Color online). Normalized density of states for both
  electron and hole in BLG. The solid and dashed lines are for the DOS
  in inhomogeneous and homogeneous systems, respectively. The
  electron (hole) band tail locates at $E<0$ ($E>0$), which gives rise
  to electron (hole) puddles at $E<0$ and $E>0$.
}
\label{fig:bdos}
\end{figure}

In the presence of potential fluctuations the density of states
(DOS) for disordered BLG is given by
$D_e(E) =  \int_{-\infty}^{E}D_0P(V)dV
     =  {D_0}{\rm erfc}(-E/\sqrt{2}s)/2$,
where $D_0 = {g_sg_v m}/(2\pi \hbar^2)$ is
the DOS in a homogeneous BLG system,
where $g_s=2$ and $g_v=2$ are the spin and
valley degeneracies, respectively.
We have $D_0=2.8\times 10^{10}$ cm$^{-2}$/meV assuming $m=0.033m_e$. The DOS
of hole can be calculated from the following relation: $D_h(E) = D_e(-E)$. In
Fig. \ref{fig:bdos}, the density of states of both electron and hole
are shown for the inhomogeneous BLG system. In the presence of
potential fluctuations, the electron and hole coexist for certain
amount of regions near CNP and their DOS approach to the homogeneous
case as the carrier energy further increases.

Because BLG is also a gapless semiconductor like MLG,  the direct thermal excitation from
valence band to conduction band at finite temperatures composes
an important source of temperature dependent transport in BLG.
Thus, the temperature dependence of thermally
excited electron density is first to be considered.

\subsection{$n_e(T)$ of BLG at CNP ($E_F = 0$)}

\begin{figure}
\epsfysize=1.8in
\epsffile{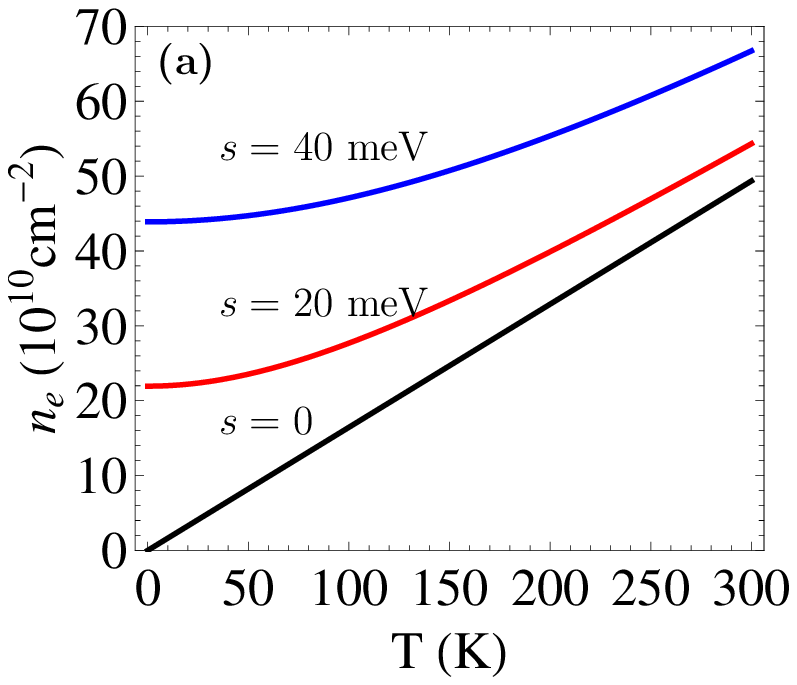}
\epsfysize=1.8in
\epsffile{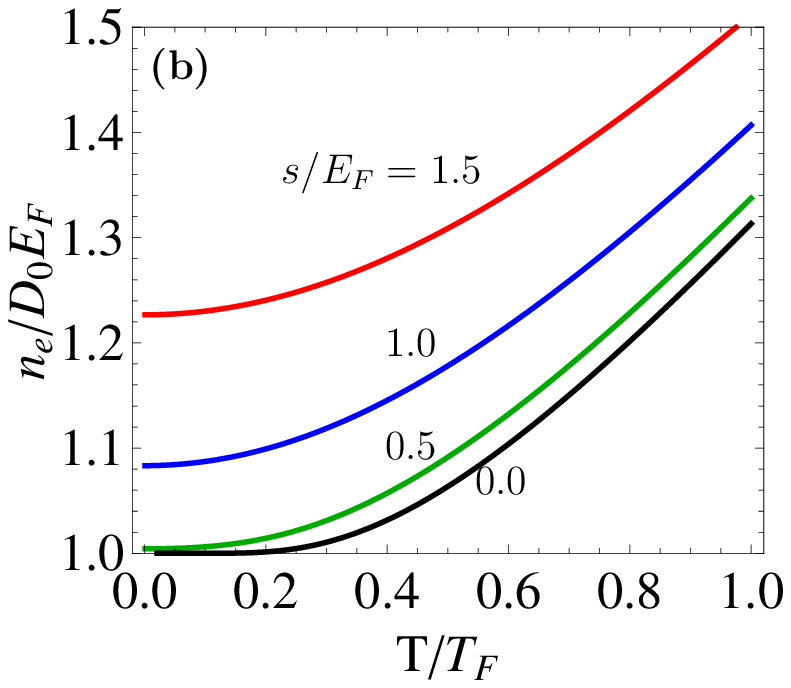}
\epsfysize=1.8in
\epsffile{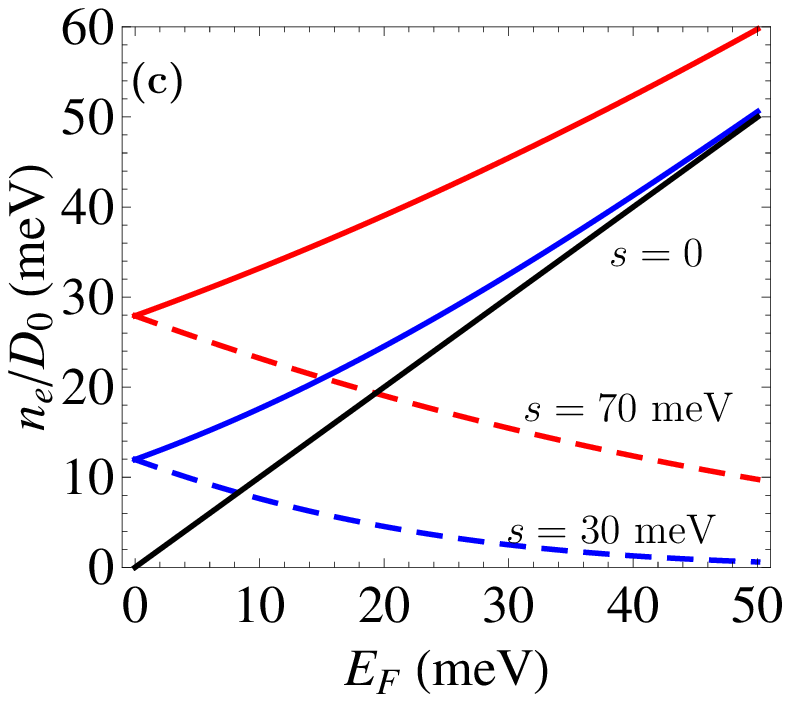}
\caption{ (Color online)
(a) The electron density of BLG at CNP as a function of
  temperature for different   $s$. At $T=0$ the density is given by
  $n_0=D_0 s/\sqrt{2\pi}$.
(b) The temperature dependent electron density of BLG at finite $E_F$
for different $s$. For $s/E_F \neq 0 $
  the leading order behavior is quadratic while at $s=0$ the density
  is exponentially suppressed.
(c) Total electron densities (solid lines) and hole densities (dashed
lines) of BLG  as a function of $E_F$ for two
different $s=30$ meV and 70 meV. The linear line represents the density
difference $n=n_e-n_h=D_0E_F$, which linearly depends on the Fermi
energy. The densities at the band tails are given by $n_e(E_F=0)=n_h(E_F=0)=D_0
s/\sqrt{2\pi}$.
\label{fig:bden}
}
\end{figure}

With the help of Eq. \ref{eq:mdenint}, we could get the total electron
density for BLG in the presence of electron-hole puddles. We first
consider the situation at CNP, where all
electrons are located in the band tail at $T=0$ and
the electron density in the band tail  is given by
$n_0=n_e(E_F=0)= {D_0 s}/{\sqrt{2\pi}}$\cite{David_PRB10}.
Contrast to the quadratic dependence of $s$ in MLG, the electron
density in the band tail for BLG is linearly
proportional to the standard deviation $s$. Unlike MLG, which has the
exact formula for $n_0(T)$ (i.e., Eq.~(\ref{eq:den_0})), we could only find
the asymptotic behavior of $n_0(T)$ at finite temperatures for BLG.
The low temperature ($k_BT/s \ll 1$) behavior of electron density at
CNP becomes
\begin{equation}
n_e(T) = n_0 \left [ 1 + \frac{\pi^2}{6} \left ( \frac{k_BT}{s} \right
  )^2 \right ].
\label{eq:bden_0}
\end{equation}
Thus, the electron density increases quadratically at low temperature
limit. For homogeneous BLG with the constant DOS
the electron density at finite temperatures is
given by
$n_e(T)=D_0\ln(2)k_BT$, which has the universal slope $D_0 \ln(2)k_B$.
The presence of the band tail suppresses the thermal excitation of
electrons and gives rise to the quadratic behavior. However,
at high temperature limit,  the density increases linearly with the same slope approaching to the homogeneous system, i.e.,
\begin{equation}
n(T) \sim D_0 \left [ \ln(2) k_BT + \frac{1}{8}\frac{s^2}{(k_BT)^2}
  \right ].
\label{eq:bden_0h}
\end{equation}
In Fig.~\ref{fig:bden}(a) we show the temperature dependent electron
density at CNP for different standard deviations. Compared with the inset of Fig.~\ref{fig:sig_mu0}, it is apparent that, even for the same strength of potential fluctuation $s$, the effects of thermal excitation of carrier density are much stronger in BLG than in MLG sample, which leads to more easily observed insulating behavior in BLG samples.

\subsection{$n_e(T)$ of BLG at finite doping ($E_F > 0$)}

In this subsection, we derived the total electron density at finite
temperatures for inhomogeneous BLG away from CNP. Contrary to MLG, we
need to calculate the finite temperature chemical potential (i.e.,
Eqs. \ref{eq:mhomochem} and \ref{eq:mchem}). The charge conservation
relation in both homogeneous and inhomogeneous BLG gives the
temperature independent chemical potential $\mu \equiv E_F$, allowing
us to directly calculate the total effective electron (hole)
density. In the case of finite gate voltage, i.e., $E_F\neq 0$, the
electron density of the homogeneous BLG for $s=0$ is given by
\begin{equation}
n_{0e}(T)=D_0E_F \left [1+ t \ln \left (1+e^{-1/t}
  \right ) \right ],
\end{equation}
where $t=T/T_F$ and $T_F = E_F/k_B$. The thermal excitation
is exponentially suppressed due to the Fermi function at low
temperatures ($T \ll T_F$). While
at high temperatures ($T \gg T_F$) it increases linearly.
In the presence of finite potential fluctuations ($s \neq 0$),  the electron and hole density at zero temperature for the inhomogeneous system are given by:
\begin{equation}
\begin{array}{l l l }
n_e(0) = {D_0 E_F} \left [ \dfrac{1}{2}{\rm erfc} \left(
  \dfrac{-1}{\sqrt{2}\tilde{s}}
  \right ) + \dfrac{\tilde{s}}{\sqrt{2\pi}} e^{-1/2\tilde{s}^2} \right ],\\
  \\
  n_h(0) = {D_0 E_F} \left [- \dfrac{1}{2}{\rm erfc} \left(
  \dfrac{1}{\sqrt{2}\tilde{s}}
  \right ) + \dfrac{\tilde{s}}{\sqrt{2\pi}} e^{-1/2\tilde{s}^2} \right ]
  \end{array}
\end{equation}
where $\tilde{s}=s/E_F$ and the difference of electron and hole density ($n = n_e - n_h = D_0 E_F$) is independent of the strength of potential fluctuation $s$ (see Fig. \ref{fig:bden}(c)).  At low  temperatures ($T \ll T_F$) the
asymptotic
behavior of the electron density is given by
\begin{equation}
n_e(T) = n_e(0) + D_0 E_F  \frac{\pi^2
}{12\sqrt{2}}\frac{e^{-1/2\tilde{s}^2}}{\tilde{s}}
\left ( \frac{T}{T_F} \right )^2.
\label{eq:bden_mu}
\end{equation}
The leading order quadratic behavior of $n_e(T)$ as in
undoped BLG ($E_F=0$) is strongly
suppressed by potential fluctuation. For the situation $s > E_F$,
the existence of electron-hole puddles gives rise to a notable
quadratic behavior [see Fig.~\ref{fig:bden}(b)]. At high temperatures
  ($T \gg T_F$)  we find
\begin{equation}
n_e(T)= n_{0e}(T)  +\frac{D_0E_F}{(1+e^{\beta E_F})^2}
\frac{\tilde{s}^2}{2} \frac{T_F}{T}.
\end{equation}
where the linear temperature dependence of electron density is dominant as the homogeneous system.

\section{conductivity of inhomogeneous BLG}
\label{sec:bcon}
With the help of total electron and hole density calculated above, we will derive the temperature dependent conductivity for BLG in the presence of electron-hole puddles. We will apply both Boltzmann theory\cite{DasEnrico_PRB10} and effective medium theory\cite{kirkpatrick1973} to interpret the intriguingly insulating behavior observed in BLG samples\cite{zhu2009, feldman2009,Nam_PRB10}.

The density and temperature dependent average conductivities in BLG, denote as $\sigma_e$ and $\sigma_h$, are given within the Boltzmann transport theory:
\begin{equation}
\begin{array}{l l l }
\sigma_e = \dfrac{n_e e^2 \langle \tau \rangle}{m}
\\
\\
\sigma_h = \dfrac{n_h e^2 \langle \tau \rangle}{m}
\label{eq:bsigeh}
\end{array}
\end{equation}
where $n_e$ and $n_h$ are average electron and hole densities,
respectively. $\langle \tau \rangle$ is the transport relaxation time for bilayer graphene:
\begin{equation}
\langle \tau \rangle = \dfrac{\int d\epsilon D_e(\epsilon) \epsilon \tau(\epsilon)(-\partial f/\partial\epsilon)}{\int d\epsilon D_e(\epsilon) f(\epsilon)}
\label{eq:bavtau}
\end{equation}
and $\tau(\epsilon)$ is calculated with Eq. \ref{eq:mscatt}. But
for BLG systems, one needs to use the parabolic dispersion relation
$\epsilon_{p\mathbf{k}} = p \hbar^2 k^2/2m$ for the pseudo-spin state
``$p$" and the static dielectric screening function derived in
Ref.~[\onlinecite{EH_PRL08}]. The wave function form factor associated
with the chiral nature of BLG is also different from the case in MLG,
which is given by $g(\theta_{\bf kk'})=\left[1+\cos2\theta_{\bf
    kk'}\right]/2$. To determine the average scattering time in BLG,
we take into account the long-range charged impurity scattering and
short-range defect scattering, which has been established that both
contribute significantly to bilayer graphene transport
properties\cite{DasEnrico_PRB10}.  The activated
conductivities should also be included in the presence of
density inhomogeneity in the BLG, which follow the same relation as given for
MLG :
\begin{subequations}
\begin{eqnarray}
\sigma_e^{(a)}(V) & = & \sigma_e \exp[\beta (E_F-V)], \\
\sigma_h^{(a)}(V) & = &\sigma_h \exp[\beta (V-E_F)],
\end{eqnarray}
\end{subequations}

\subsection{$\sigma(T)$ of BLG at CNP}
\begin{figure}
\includegraphics[width=6.5cm]{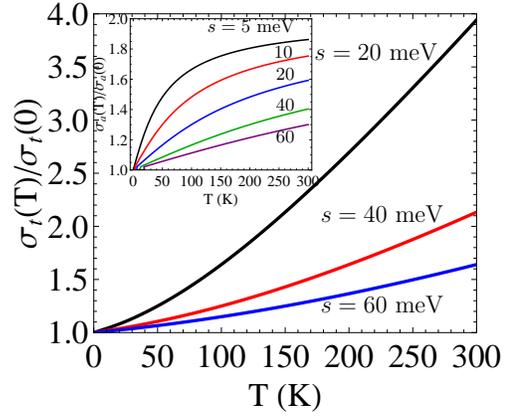}
\caption{(Color online)
$\sigma_t(T)$ of BLG at charge neutral point for
different $s$ (Eq.~\ref{eq:msig1} with $n_e$ for BLG). Inset shows the thermally activated conductivity in BLG as a
function of temperature, where $\sigma_a(T)/\sigma_a(0)=1+ e^{\beta^2 s^2/2}
  {\rm erfc} (\beta s/\sqrt{2}) $, the same as for the MLG case.
\label{fig:bsig_mu0}
}
\end{figure}

When electron-hole puddles form in the BLG samples (denote the electron (hole) puddle as region `1' (`2')), the transport properties can be treated with effective medium theory as described in Sec. \ref{sec:mono}. And Eqs. \ref{eq:sig1}-\ref{eq:siglast} for the inhomogeneous MLG also apply to the inhomogeneous BLG system. We will first discuss the total conductivity of BLG at CNP ($E_F =0 $). In this case, the electron and hole are equally occupied and the total conductivity $\sigma_t = \sigma_1$ (see Eq. \ref{eq:msig1} and \ref{eq:siglast}). At low temperature limit ($T \ll s/k_B$), the activated conductivities increase linearly with a slope
$\sqrt{2/\pi}k_B/s$ as the temperature increases. The next order temperature correction to the conductivity is quadratic $T^2$, which arises from the thermal activation (see Eq. \ref{eq:bden_0}). Thus, at low temperature limit the total conductivity at
CNP is given by
\begin{equation}
\sigma_t(T) = \sigma(0) \left [1+ \sqrt{\frac{2}{\pi}}\frac{k_BT}{s} +
  \frac{\pi^2}{6} \left ( \frac{k_BT}{s} \right )^2 \right ].
\end{equation}
At high temperatures ($k_BT \gg s$), the total conductivity is given by:
\begin{equation}
\sigma_t = \sigma_e \left [ 2 - \sqrt{ \frac{2}{\pi}} \frac{s} {k_BT} +
  \frac{s^2}{2 (k_BT)^2} \right ].
\end{equation}
It is apparent that the activation behavior approaches
a limiting value at high temperature limit ($T \gg s/k_B$) while the thermally activated electron density becomes dominant, which increases linearly with a universal slope $\ln(2)$ regardless of
the sample quality. Thus,  all temperature dependence of the total conductivity comes from the thermal
excitation through the change of the carrier density given in
Eq.~(\ref{eq:bden_0h}). In Fig.~\ref{fig:bsig_mu0} we show the calculated
temperature dependent conductivity at charge neutral point. The inset
present the activated conductivity versus the temperature. In
  Ref. [\onlinecite{zhu2009}], $\sigma_t (T)$ at the CNP of the BLG
  sample increases almost two times as temperature $T$
varies from 4 K to 300 K. Our theoretical analysis using a potential fluctuation parameter
$s\sim 40 $ meV gives reasonable agreement with the experimental data.

\subsection{$\sigma(T)$ of BLG at finite doping ($E_F > 0$)}
The temperature dependent conductivities at finite doping ($E_F > 0$)
are very complex because three energies ($E_F$, $s$, and
$k_BT$) are competing. Regardless of
$E_F$,  when $k_BT \ll s$, we have the asymptotic behavior of
conductivities in region 1
and 2 the same as MLG situation, given in Eq.~(\ref{eq:msigt}). But the average electron and hole conductivities ($\sigma_e$ and $\sigma_h$) are quite different from MLG case, which is determined by the specific band dispersion relation and also the dielectric function $\epsilon(q,T)$. Thus, the leading order correction to $\sigma_t$ in BLG is also
linear, which comes from the activated conductivity, but the coefficient is exponentially suppressed by the term
$\exp(-E_F^2/2s^2)$. In the high mobility sample
with small $s$, the activated conductivity is weakly temperature
dependent except around CNP, i.e. $E_F < s$.
Since the density increase by thermal excitation is also suppressed
exponentially by the same factor [see Eq.~(\ref{eq:bden_mu})] the
dominant temperature dependent conductivity arises from the scattering
mechanism\cite{dassarma2010}. On the other hand,  in the low mobility sample with large value of $s$, the linear temperature dependence due to thermal activation can be observed even at high
carrier densities $E_F \agt s$.

\begin{figure}[tb]
 \begin{center}
  \includegraphics[width=6.5cm]{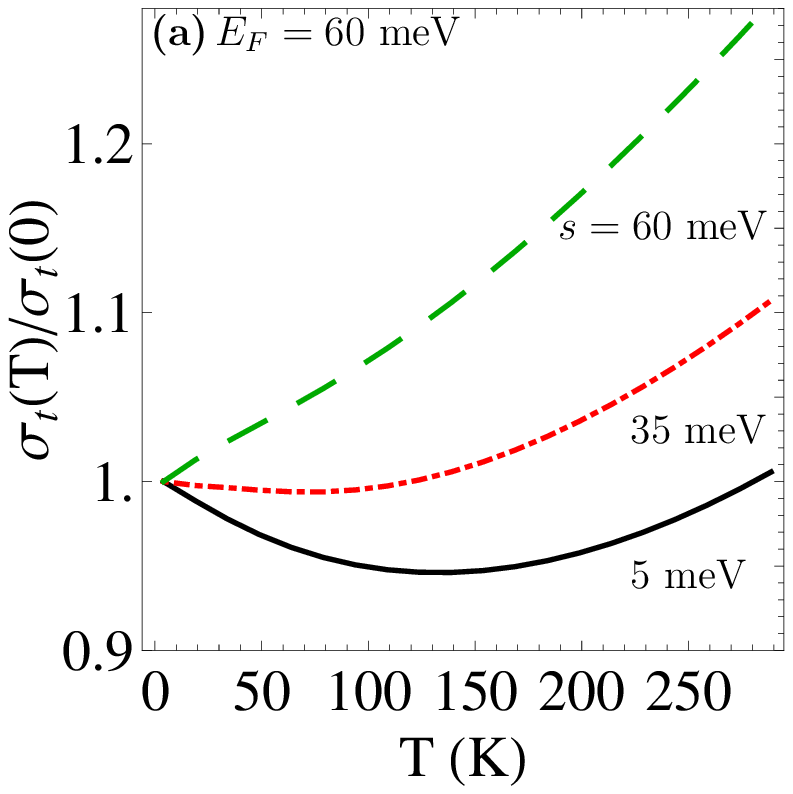}
  \includegraphics[width=6.5cm]{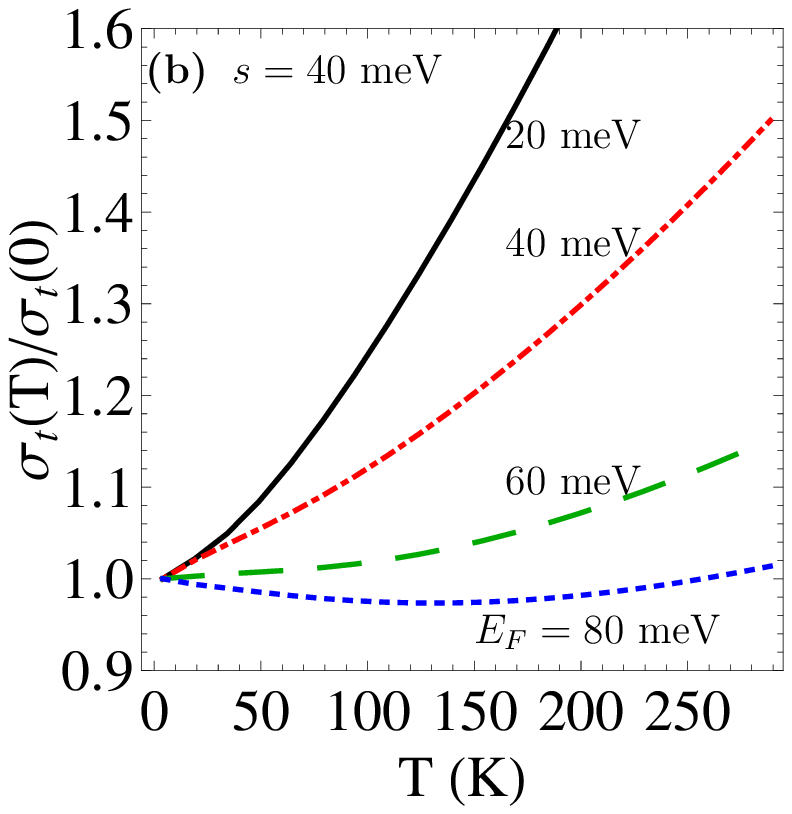}
  \caption{
           (Color online).  Calculated total conductivity $\sigma_t(T)/\sigma_t(0)$ of BLG with the following parameters: $n_i = 10^{12}$ cm$^{-2}$ and $n_d V_0^2 = 2$ (eV {\AA})$^2$.
           (a) $\sigma_t(T)$ for  $E_F=60$ meV and for
    different $s$.
           (b) $\sigma_t(T)$
for $s=40$ meV and for
    several $E_F=20$, 40, 60, 80 meV, which correspond to the net carrier densities $n=n_e-n_h=0.55\times 10^{12}$, $1.1\times 10^{12}$, $1.6\times 10^{12}$, and $2.2\times 10^{12}$ cm$^{-2}$.
          }
  \label{fig:bsig_tot}
 \end{center}
\end{figure}
In Fig.~\ref{fig:bsig_tot} we calculate the total conductivities
(a) for a fixed $E_F$ and several $s$ and (b)
for a fixed $s$ and several $E_F$.  Even for homogeneous BLG, there
are two scattering mechanism competing with each other. The
short-range disorder in BLG contributes to a strong insulating
transport behavior for all temperature, whereas screened Coulomb
scattering always leads to a metallic behavior for $T \ll
T_F$\cite{DasEnrico_PRB10}. At low temperature limit, the total
conductivity $\sigma_t(T)$ decreases with increasing temperature,  but
at higher temperatures, the short-range disorder contribution becomes
quite big and leads to a $\sigma_t(T)$ increasing with $T$. Therefore,
when $s$ is small, the scattering mechanism is dominant and the total
conductivity manifests a non-monotonic temperature dependence (see
Fig. \ref{fig:bsig_tot}(a)).  However, for large $s$ the
activated temperature dependence
behavior overwhelms the metallic temperature dependence,
and the system shows
insulating behavior (see Fig. \ref{fig:bsig_tot}(b)). It clearly shows
that the insulating behavior in BLG sample appeared at carrier
densities as high as $10^{12}$ cm$^{-2}$ or higher.

\section{Connection to earlier theories}
\label{sec:connection}

We have demonstrated theoretically that the observed insulating
behavior in temperature-dependent monolayer and bilayer graphene
conductivity can be explained by the thermal activation between puddles.
There are also other theories which have been elaborated to
explain low carrier density graphene transport
\cite{Adam_PNAS07,RossiAdamDas_PRB09,DasEnrico_PRB10}. In this section, we establish the bridge
to connect our current theory and earlier theories on graphene
transport due to the formation of inhomogeneous electron-hole puddles
near the charge neutrality point.

The key qualitative difference between our theory and all earlier graphene transport theories is the introduction of the 2-component transport model where regular diffusive metallic carrier transport coexists with local activated transport due to activation across potential fluctuations in the puddles.  Our theory just explicitly accounts for the inhomogeneous landscape in the system, which earlier theories ignored.  This 2-component nature of graphene transport, where both metallic and insulating behavior coexist because of the existence of puddles, produces the experimentally observed complex temperature dependence with the low-density behavior being primarily insulating-like and the high-density behavior being primarily metallic-like.

Two different theories have been developed  to study the low-density
transport in graphene, where the strong density inhomogeneity is
dominated. In Ref.~[\onlinecite{Adam_PNAS07}] Adam {\it et al.}
qualitatively  explained the
plateau-like approximate nonuniversal minimum conductivity at low
carrier density observed in monolayer graphene samples.
The basic idea is to introduce an approximate pinning of the carrier
density at $n=n^*\approx n_i$ at low carrier density limits
$|n|<|n_i|$, where $n_i$ is an impurity density.
The constant minimum
conductivity is then given by $\sigma_{min} \sim \sigma(n=n_i)$ for $n<n_i$.
This simple theory for monolayer graphene
transport qualitatively explained the existence of conductivity
minimum plateau and the extent to which the minimum conductivity
is not universal, which was in good agreement with the observed
density-dependent conductivity over a wide range of charged impurity
densities\cite{TanDas_PRL07,Chen2008}. However, this theory did not
take account of the highly heterogeneous structure near charge
neutrality point and the thermally activated conductivity at finite
temperatures, which then can not explain the observed non-monotonic
temperature dependent transport in low mobility graphene
samples\cite{Heo_arXiv10}.

A more elaborate Thomas-Fermi-Dirac (TFD) theory and an effective
medium approximation (EMT) have been introduced in
Refs.~[\onlinecite{Rossi_PRL08}] and [\onlinecite{RossiAdamDas_PRB09}] to
study the electrical transport properties  of disordered monolayer
graphene. The ground state-density landscape $n({\bf r})$ can be
obtained within this TFD approach and the resultant electrical
transport can be calculated by averaging over disorder realizations
and the effective medium theory. This theory gives a finite
minimum conductivity and is able to explain the crossover of the
density-dependent conductivity from the minimum value at the Dirac
point to its linear behavior at higher doping. Later, this TFD-EMT
theory is also applied to calculate the conductivity of disordered
bilayer graphene in Ref.~[\onlinecite{DasEnrico_PRB10}]. The TFD-EMT technique successfully explains the graphene, both MLG and
BLG, transport properties in the theoretically difficult
inhomogeneity-dominant regime near the charge neutral point, but this
approach fails to explain the
temperature dependence of the conductivity for a wide range of
temperatures.

In our current model discussed above, we include three effects, the
electron-hole structure formation, the thermal activated
conductivities and the temperature dependence of screening effects, to
explain the temperature-dependent conductivity in both monolayer and
bilayer graphene systems. The nonmonotonic temperature-dependent
conductivity in graphene systems is then naturally understood from the
competition between the thermal activation of charge carriers and the temperature-dependent screening effects. Our transport theory
qualitatively explains the observed coexisting metallic and insulating
transport behavior in both MLG and BLG systems. For low mobility MLG
samples, the dominant role on graphene conductivity switches from the
thermally activated transport of inhomogeneous electron-hole puddles
to metallic temperature-dependent screening effects, which
gives rise to a nonmonotonic behavior from the strong insulating
behavior at low temperatures to metallic
behavior at high temperatures. On the other hand, another nonmonotonic
temperature-dependent transport can be observed in very high mobility
bilayer graphene devices, i.e., from metallic behavior at low
temperatures due to the screening effects of Coulomb scattering to insulating behavior
at high temperatures due to the short-range disorder. The merit of
our model is that it is so simple that we could get the asymptotic
behavior at low and high temperature limits analytically. Moreover, it
provides a clear physical picture of the dominant mechanisms at
different regimes as discussed above.

\section{Discussions and Conclusions}
\label{sec:conclu}

We first discuss the similarity and the difference
between MLG and BLG transport from the perspective of our
transport-theory considerations. We find that both manifest an
insulating behavior in $\sigma_t(T)$ for low mobility samples. We also
find that both systems could exhibit a non-monotonic temperature
dependent conductivity for low mobility samples.
However, the physical origin for the
non-monotonic temperature dependence is quite different in the two
systems: in the MLG the non-monotonic feature comes from the
competition between thermal activation and the metallic screening
effects, which leads to $\sigma_t(T)$ first increasing and then
decreasing with increasing temperature (see
Fig. \ref{fig:sig_act}(a)). While for BLG, the competition between
short-range insulating  scattering and metallic Coulomb screening
effects leads to $\sigma_t(T)$ first decreasing and then increasing as
temperature increases (see Fig. \ref{fig:bsig_tot}(a)). Most important
quantitative difference between MLG and BLG transport comes from their
band dispersions, which leads to much weaker effects of density inhomogeneity
in MLG so that the anomalous insulating temperature dependence of
$\sigma(T)$  is typically not observed in MLG away from the CNP
although the gate voltage dependence of MLG and BLG conductivities are
similar\cite{Morozov_RPL08,Shudong_PRB10}. The linear Dirac carrier
system for MLG leads to linear DOS, which goes to zero at CNP, but the
parabolic band dispersion relation in BLG leads to a constant DOS. Due to
the difference in the density of states between homogeneous MLG and
BLG, the modified DOS in
inhomogeneous MLG is increased (see Fig. \ref{fig:mdos}) rather than decreased
in inhomogeneous BLG (see Fig. \ref{fig:bdos}).
The dimensionless potential fluctuation strength
$\tilde s$ ($\equiv s/E_F$) is much weaker in MLG than in BLG from
simple estimates: $\tilde s_{BLG}/ \tilde s_{MLG} \sim 32/ \sqrt{\tilde
  n}$
where $\tilde n = n/10^{10}$, and $\tilde s_{BLG} \gg \tilde
s_{MLG}$ upto $n=10^{13}$ cm$^{-2}$.
Direct calculations \cite{dassarma2010} show that the self-consistent
values of $s$ tend to be much larger in BLG than in MLG for identical
impurity disorder. In addition, the qualitatively different DOS leads
to much stronger effective short-range scattering in BLG compared with
MLG even for the same bare scattering strength. Thus, the insulating
behavior in $\sigma_t(T)$ will show up at high temperatures even for
relatively higher mobility BLG samples (i.e., small $s$). In contrast,
only in very low mobility MLG samples,
where $s$ is very large,
can the insulating behavior of temperature dependent resistivity  be
observed\cite{YTan_EPJT07,Heo_arXiv10}. No simple picture would apply
to a gapped ($\Delta_g$) BLG system, since four distinct energy scales
($s, E_F, k_B T$, and $\Delta_g$) will compete and the conceivable
temperature dependence depends on their relative
values\cite{castro2007,oostinga2008,mak2009}. Our
assumption of BLG quadratic band dispersion is valid only at low
($\lesssim 5 \times 10^{12}$ cm$^{-2}$) carrier densities, where most
of the current transport experiments are carried out. At higher
densities the band dispersion is effectively linear and the disorder
effects on $\sigma_t(T)$ are weaker.

Before concluding, we emphasize that our theory is physically motivated since puddles are experimental facts in all graphene samples.  Puddles automatically imply a 2-component nature of transport since both diffusive carriers and activated carriers can, in principle, contribute to transport in the presence of puddles.  Of course, the effect of puddles is much stronger at low carrier densities, explaining why insulating (metallic) temperature dependence is more generic at low (high) graphene carrier densities.  We emphasize that local carrier activation in puddles is just one of (at least) four different independent transport mechanisms contributing to the temperature dependent conductivity.  The other three are temperature dependent screening (Ref.~[\onlinecite{HwangScreen_PRB09}]), phonons (Refs.~[\onlinecite{HwangDasPhonon_PRB08,MinHwang_PRB11}]), and Fermi surface thermal averaging (Refs.~[\onlinecite{HwangScreen_PRB09,Muller_PRL09}]).  Our theory presented here includes the three electronic mechanisms for temperature dependence: screening, Fermi surface averaging, and puddle activation.  We leave out phonons, which have been considered elsewhere (Ref.~[\onlinecite{HwangDasPhonon_PRB08,MinHwang_PRB11}]) and will simply add to the temperature dependent resistivity.  The weak phonon contribution to graphene resistivity makes it possible for the electronic mechanisms to dominate even at room temperatures, but obviously at high enough temperatures, the system will, except perhaps at the lowest densities around the CNP,  manifest metallic temperature dependence with the resistivity increasing with temperature because of phonon scattering.  Similarly, the puddle effects dominate low densities and therefore, the insulating behavior will persist to very high temperatures around the zero-density CNP since activation across potential fluctuations are dominant at the CNP.  It is gratifying to note that these are precisely the experimental observations. We note that in general the temperature dependent conductivity of graphene could be very complex since many distinct mechanisms could in principle contribute to the temperature dependence depending on the carrier density, temperature range, and disorder in the system.  Inclusion of phonons (at high temperatures) and quantum localization (at low temperatures) effects, which are both neglected in our theory, can only complicate things further.  What we have shown in this work is that the low-density conductivity near the CNP is preferentially dominated by density inhomogeneity and thermal carrier activation effects leading to an insulating temperature dependence in the conductivity whereas the high-density conductivity, where the puddles are screened out, is dominated by a metallic conductivity due to temperature-dependent screening effects. This general conclusion is consistent with all experimental observations in both MLG and BLG systems to the best of our knowledge except at very high temperatures where phonon effects  would eventually lead to metallic behavior at all densities.

To conclude, we have investigated both MLG and BLG transport in the
presence of electron-hole puddles within an analytic statistical theory.
Our theory explains the experimentally measured insulating behavior at
low temperatures and the
consequent nonmonotonic behavior for low
mobility samples \cite{Heo_arXiv10,Nam_PRB10,zhujun_PRBR10}. A reasonable quantitative agreement with the experimental data can be obtained by choosing appropriate disorder parameters in our theory (i.e. potential fluctuation and impurity  strength) for different samples. We find that the puddle parameter $s$, defining typical potential fluctuations, to be around $10-80$ meV in typical graphene samples as extracted by fitting our theory to existing experimental transport data near the charge neutrality point.  These values of potential fluctuations characterizing the graphene charge neutrality point are very consistent with direct numerical calculations of graphene electronic structure in the presence of quenched charged impurities\cite{dassarma2010,RossiAdamDas_PRB09,Rossi_PRL08,DasEnrico_PRB10}. We also relate our current model to
earlier theories using the picture of diffusive transport through
disorder-induced electron-hole puddles.
Finally, we show the
similarity and the quantitative difference between MLG and BLG
transport in the presence of puddles.

\begin{acknowledgments}

QL acknowledges helpful discussions with
D. S. L. Abergel. The work is supported by ONR-MURI, NRI-NSF-SWAN.

\end{acknowledgments}

\appendix
\section{A self-consistent formulation of graphene density of states
  in the presence of inhomogeneity}
\label{sec:appen}

Below we provide a microscopic theory to calculate self-consistently
the electronic density of states in the presence of the
potential fluctuations caused by random charged
impurities located near graphene/substrate interface,
which has been applied to two dimensional semiconductor
based electron gas systems\cite{Stern_SS76}. This self-consistent
approach mainly addresses two problems with the presence of random
charged impurities. One is the screening of the long-range Coulomb
interactions between the carriers  and the charged impurities. The
other is the real-space potential fluctuations produced by the random
array of charged impurities.

The motivation for this appendix is two-fold: (1) providing a microscopic self-consistent theory of graphene density of states in the presence of puddles; (2) showing that our approximate physically-motivated density of states (Eq.~\ref{eq:mdos}) is an excellent approximation to the self-consistent density of states.

\subsection{Monolayer graphene}

First, we apply the self-consistent consideration of random charged
impurities on the density of states in monolayer graphene.

\begin{figure}[tb]
\includegraphics[width=0.8\linewidth]{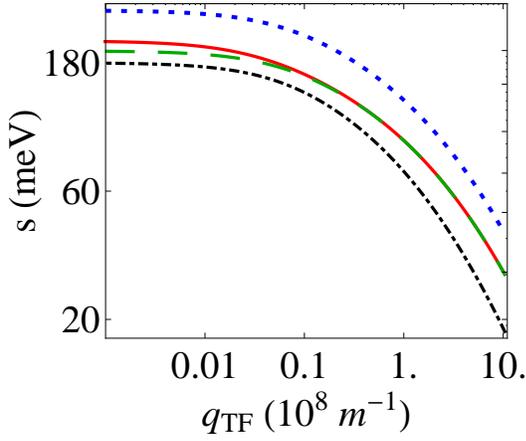}
\caption{ (Color online)
 Standard deviation of potential fluctuation $s$ versus the screening constant $q_{TF}$  (loglog plot) in MLG by varying the Fermi level.  The dotted blue line is for $n_{imp}=1.0 \times 10^{12}$ cm$^{-2}$, $z_0 = 1$ nm and $d=100$ nm. The solid red line is for $n_{imp}=0.5 \times 10^{12}$ cm$^{-2}$, $z_0 = 1$ nm and $d=200$ nm. The dashed green line is for $n_{imp}=0.5 \times 10^{12}$ cm$^{-2}$, $z_0 = 1$ nm and $d=100$ nm.  The dotdashed black line is for $n_{imp}=0.5 \times 10^{12}$ cm$^{-2}$, $z_0 =2$ nm and $d=100$ nm.
}
\label{fig:parameter}
\end{figure}

\begin{figure}
\includegraphics[width=0.75\linewidth]{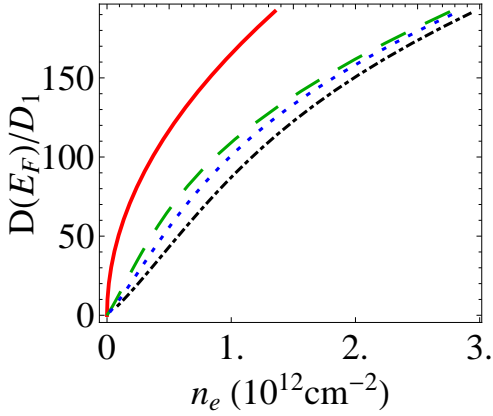}
\caption{ (Color online)
 Calculated the density of states of electron $D_e(E_F)$ in MLG versus electron density using the following parameters: the insulator thickness $d=100$ nm, the impurity distance from the interface $z_0=1$ nm. The solid red line is for unperturbed density of states. The dashed green, dotted blue and dotdashed black lines are corresponding to  $n_{imp} = 0.5$, $1.0$ and $2.0\times 10^{12}$ cm$^{-2}$, respectively.
}
\label{fig:mdosden}
\end{figure}

The simple theory of linear screening gives\cite{Stern_SS76}:
\begin{equation}
q_{TF}=\dfrac{2 \pi e^2}{\kappa}D_e(E_F)
\label{eq:s}
\end{equation}
where $q_{TF}$ is the Thomas-Fermi screening wavevector, $D_e(E_F)$ is
the density of states at the Fermi level and $\kappa$ is the dielectric
constant ($\kappa \simeq 2.5$ for graphene on SiO$_2$ substrate).

The screening constant shown in Eq. \ref{eq:s} enters Poisson's
equation for the potential change $\phi(r,z)$ produced by a charge
density $\rho_{ext}$ (associated with the charged impurities in the
graphene/substrate environment). For a charge $Z e$ (we use $Z\equiv 1$
in the calculation) located at $r \equiv \sqrt{x^2+y^2}=0$ and
$z=z_0>0$ (on top of the graphene layer), the additional Coulomb potential
satisfies:
\begin{equation}
\nabla^2\phi(r,z)-2 q_{TF}g(z)\phi_0(r)=-\dfrac{4\pi Z e \delta(x) \delta(y)\delta(z-z_0) }{\kappa_0}
\label{eq:poisson}
\end{equation}
where $\kappa_0= \kappa_v = 1.0$ in the vacuum ($z>0$),
$\kappa_0= \kappa_{ins}= 3.9$ in SiO$_2$ ($z<0$) and $\kappa=\frac{\kappa_{ins}+\kappa_v}{2}$. For graphene,
$g(z)=\delta(z)$ is the carrier density distribution normal to the
interface and $\phi_0(r)=\int \phi(r,z) g(z) dz = \phi(r,0)$.

To solve Eq. \ref{eq:poisson} we take advantage of the cylindrical symmetry to write\cite{Stern_PR67} :
\begin{equation}
\phi(r,z)=\int_0^{\infty}J_0(k'r)A_{k'}(z)k'dk'
\end{equation}
The potential will satisfy Eq. \ref{eq:poisson} if
\begin{equation}
\dfrac{d^2A_k}{dz^2}-k^2 A_k-2 q_{TF} A_k(0)g(z)=-\dfrac{2Z e \delta(z-z_0)}{\kappa_0}
\label{eq:Ak}
\end{equation}
At the interface $z=0$, $A_k(z)$ must be continuous and satisfy
$\kappa_v (d A_k/dz) - \kappa_{ins} (d A_k/dz)=2 q_{TF} A_k(0) \kappa$. $A_k(z)$
should also satisfies the boundary condition $A_k (z)\rightarrow 0$ as
$z  \rightarrow \infty$. In addition, the impurity potential $\phi$
will go to zero at the metallic contact below the SiO$_2$ (i.e., $A_k(-d)
\equiv 0$ and $d$ is the thickness of the SiO$_2$ layer). Such
screening effects are absent in the SiO$_2$. After some
algebra, the explicit expression of $A_k(k, 0)$ for insulator
thickness $d$ and the impurity distance from graphene/substrate
interface $z_0$ is given by:
\begin{equation}
A_k(k, 0) = \dfrac{2 e^{-k z_0} Ze~\text{sinh}(d k)}{k \kappa_{ins} \text{cosh}(d k)+(k \kappa_{v}+2 q_{TF}\kappa)  \text{sinh}(d k)}
\end{equation}
For the thickness of insulator in the limit $d \rightarrow \infty$, we have
$A_k(k,0)=\dfrac{e^{-k z_0} Ze }{(k+q_{TF})\kappa}$, which has been given in the
Appendix B of Ref.~[\onlinecite{Stern_PR67}].
The potential fluctuations with an array of point charges at random
positions in the plane $z= z_0$ have a mean-square variation about the
average potential \cite{Stern_SS76}:
\begin{equation}
V_{rms}^2 = 2\pi n_{imp} e^2 \int [A_k(0)]^2 k dk
\label{eq:flucV}
\end{equation}

\begin{figure}[tb]
\includegraphics[width=0.6\linewidth]{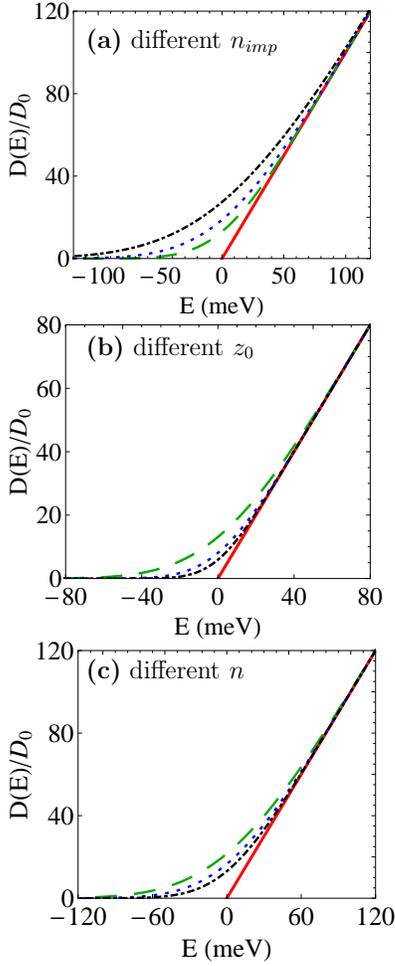}
\caption{ (Color online)
 Calculated density of states of electron $D_e(E)$ of MLG  versus energy $E$ for different impurity configurations and carrier densities $n$. The solid red lines are for the non-interacting MLG system. (a) Calculated $D_e(E)$ in MLG for the insulator thickness $d=100$ nm, the impurity distance from the interface $z_0=1$ nm and carrier density $n=2.0 \times 10^{12}$ cm$^{-2}$. The dashed green, dotted blue and dotdashed  black lines are corresponding to $n_{imp} = 0.5$, $1.0$, $2.0 \times 10^{12}$ cm$^{-2}$, respectively. (b) Calculated $D_e(E)$ in MLG for $d=100$ nm, $n_{imp} =0.5 \times 10^{12}$ cm$^{-2}$ and $n=2.0 \times 10^{12}$ cm$^{-2}$. The dashed green, dotted blue and dotdashed  black lines are corresponding to $z_0 = 1$, $2$, $3$ nm, respectively. (c) Calculated $D_e(E)$ in MLG for $d=100$ nm, $z_0=1$ nm and $n_{imp}=0.5 \times 10^{12}$ cm$^{-2}$. The dashed green, dotted blue and dotdashed  black lines are corresponding to $n = 0.5$, $1.0$, $2.0 \times 10^{12}$ cm$^{-2}$, respectively.
}
\label{fig:mdosE}
\end{figure}

To obtain specific results for the electronic density of states and
the screening constant we use the simple Gaussian broadening
approximation for the density of states\cite{Arnold_APL74}. The
disorder-induced potential energy fluctuations is described by
$P(V) = \frac{1}{\sqrt{2\pi s^2}}\text{exp}(-V^2/2 s^2)$
(Eq. \ref{eq:flucV}). Then the density of states becomes
\begin{equation}
\begin{array}{l l l }
D_e(E)= \int_{-\infty}^{E} \dfrac{g_sg_v(E-V)}{2\pi (\hbar v_F)^2} P(V)dV
\\
\\
= D_1 \big[\dfrac{E}{2} \text{erfc}(-\dfrac{E}{\sqrt{2}
    s})+\dfrac{s}{\sqrt{2 \pi}} \exp(-\dfrac{E^2}{2 s^2})\big]
\end{array}
\end{equation}
where erfc$(x)$ is the complementary error function, $s = V_{rms}$,
$D_1 =\dfrac{g_s g_v}{2\pi (\hbar v_F)^2}$, $v_F$ is the graphene
(Fermi) velocity, $g_s=2$ and $g_v=2$ are the spin and valley
degeneracies, respectively.

By choosing the chemical potential $E_F$ as a tuning parameter we
have the following coupled equations:
\begin{equation}
\begin{array}{l l l l }
q_{TF}=\dfrac{2 \pi e^2}{\kappa}D_e(E_F)
\\
s^2 = 2\pi n_{imp} e^2 \int [A_k(0)]^2 k dk
\end{array}
\end{equation}
For fixed values of $E_F$, $n_{imp}$, $d$ and $z_0$, we get the
self consistent results for $s$, $q_{TF}$ by solving the above two
coupled equations. The electron density could be gotten from the
formula:
\begin{equation}
n_e = \int_{-\infty}^{\infty}D_e(\epsilon) f(\epsilon)d\epsilon
 \end{equation}
where $f(\epsilon)$ is the Fermi-Dirac distribution function. The
electron density in the presence of disorder-induced electron-hole puddles
has been discussed in Sec.~\ref{sec:mden}, where we use the potential
fluctuation $s$ as a fixed parameter. And here we self-consistently
solve the parameter $s$ from a microscopic point of view, which is in
good agreement with the results shown in Sec.~\ref{sec:mden}.
The potential fluctuation in Eq.~\ref{eq:flucV} affects the electronic
density of states. But the fluctuations depend on the screening via
Eq.~\ref{eq:Ak} while the screening depends on the density of states
via Eq. \ref{eq:s}. Therefore, we have a coupled problem which must be
solved self-consistently.

 In Fig.~\ref{fig:parameter}, the standard deviation of the potential
 fluctuation $s$ and the screening constant $q_{TF}$ are plotted for
 different values of the Fermi level. The self-consistently solved
 parameters $(s, q_{TF})$ depend on
 the fixed charged impurity density $n_{imp}$, the SiO$_2$
 thickness $d$, the location of the fixed charged impurity $z_0$,
 and the Fermi level $E_F$ (i.e. the carrier density $n$). All
 these four effects can be understood from physical intuition. The
 reduction of the SiO$_2$ thickness weakens the potential fluctuations
 when the screening length is small even though there is a little effect for
 strong screening. As the charged impurities go away from the
 graphene layer the potential fluctuations is also reduced, while
 the potential fluctuations becomes stronger with the higher impurity
 density. Increasing
 the carrier density $n$ gives rise to the stronger screening effects,
 and leads to weaker potential fluctuations.

In Fig.~\ref{fig:mdosden}, the density of states of monolayer graphene is given with the parameters of SiO$_2$ thickness $d=100$ nm and the distance of fixed charged impurities $z_0 = 1$ nm for different
impurity densities $n_{imp}$. In Fig.~\ref{fig:mdosE}, we present the
electronic density of states for different carrier densities and
impurity configurations. The self-consistent calculation of the density
of states verifies the results presented in Sec.\ref{sec:mden} as shown
in Fig.~\ref{fig:mdos}, where we choose the potential fluctuation $s$
as an adjustable parameter. For monolayer graphene, the presence of
spatially random charged impurities increases the electronic density
of states in the whole range of energy. The corresponding hole density
of states can be obtained by changing the sign of energy
$D_h(E)=D_e(-E)$.

\subsection{Bilayer graphene}

\begin{figure}[tb]
\includegraphics[width=0.8\linewidth]{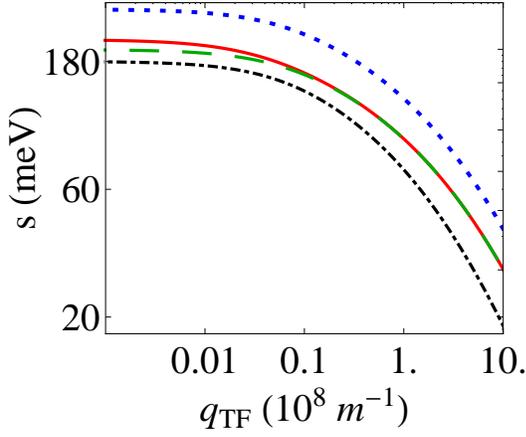}
\caption{ (Color online)
 Standard deviation of potential fluctuation $s$ versus the screening constant $q_{TF}$  (loglog plot) in BLG by varying the Fermi level.  The dotted blue line is for $n_{imp}=1.0 \times 10^{12}$ cm$^{-2}$, $z_0 = 1$ nm and $d=100$ nm. The solid red line is for $n_{imp}=0.5 \times 10^{12}$ cm$^{-2}$, $z_0 = 1$ nm and $d=200$ nm. The dashed green line is for $n_{imp}=0.5 \times 10^{12}$ cm$^{-2}$, $z_0 = 1$ nm and $d=100$ nm.  The dotdashed black line is for $n_{imp}=0.5 \times 10^{12}$ cm$^{-2}$, $z_0 =2$ nm and $d=100$ nm.
}
\label{fig:bparameter}
\end{figure}

\begin{figure}
\includegraphics[width=0.75\linewidth]{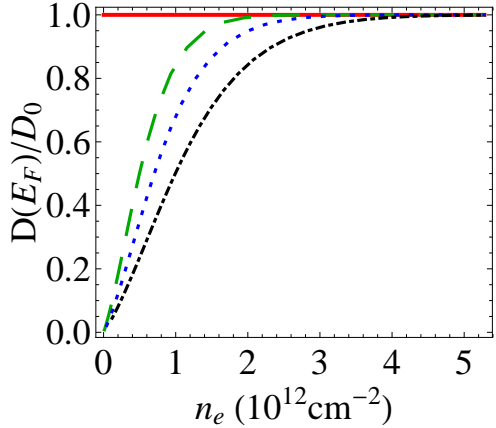}
\caption{ (Color online)
 Calculated density of states of electron $D_e(E_F)$ of BLG versus electron density using the following parameters: the insulator thickness $d=100$ nm, the impurity distance from the interface $z_0=1$ nm. The solid red line is for unperturbed density of states. The dashed green, dotted blue and dotdashed black lines are corresponding to  $n_{imp} = 0.5$, $1.0$ and $2.0\times 10^{12}$ cm$^{-2}$, respectively.
}
\label{fig:bdosden}
\end{figure}

In this subsection, we provide the
density of states in bilayer graphene in the presence of potential
fluctuations. As shown for monolayer graphene, we use
the linear screening written as\cite{Stern_SS76}:
\begin{equation}
q_{TF}=\dfrac{2 \pi e^2}{\kappa}D_e(E_F)
\label{eq:bs}
\end{equation}
where $D_e(E_F)$ is the density of states of BLG at the Fermi level and
$\kappa$ is the dielectric constant and for BLG on SiO$_2$, $\kappa
\simeq 2.5$.

\begin{figure}[tb]
\includegraphics[width=0.6\linewidth]{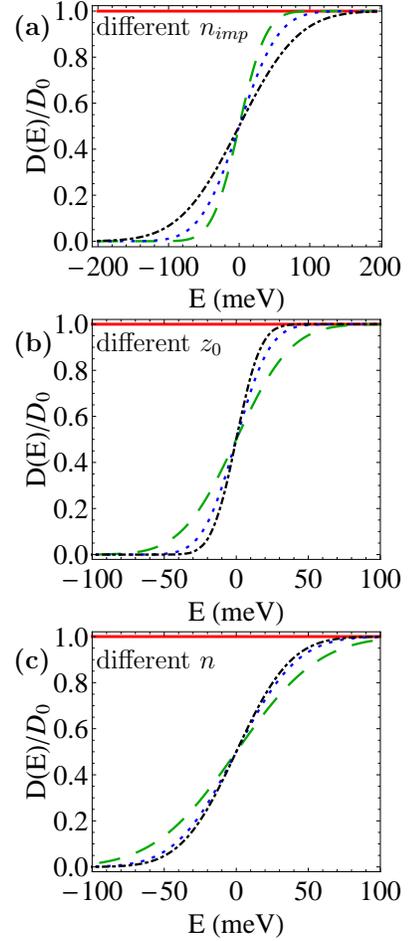}
\caption{ (Color online)
 Calculated density of states of electron $D_e(E)$ of BLG  versus energy $E$ for different impurity configuration and carrier densities $n$. The solid red lines are for the non-interacting BLG system. (a) Calculated $D_e(E)$ in BLG for the insulator thickness $d=100$ nm, the impurity distance from the interface $z_0=1$ nm and carrier density $n=2.0 \times 10^{12}$ cm$^{-2}$. The dashed green, dotted blue and dotdashed  black lines are corresponding to $n_{imp} = 0.5$, $1.0$, $2.0 \times 10^{12}$ cm$^{-2}$, respectively. (b) Calculated $D_e(E)$ in BLG for $d=100$ nm, $n_{imp} =0.5 \times 10^{12}$ cm$^{-2}$ and $n=2.0 \times 10^{12}$ cm$^{-2}$. The dashed green, dotted blue and dotdashed  black lines are corresponding to $z_0 = 1$, $2$, $3$ nm, respectively. (c) Calculated $D_e(E)$ in BLG for $d=100$ nm, $z_0=1$ nm and $n_{imp}=0.5 \times 10^{12}$ cm$^{-2}$. The dashed green, dotted blue and dotdashed  black lines are corresponding to $n = 0.5$, $1.0$, $2.0 \times 10^{12}$ cm$^{-2}$, respectively.
}
\label{fig:bdosE}
\end{figure}

Following the same procedure discussed for MLG, the disorder-induced
potential fluctuation is described by the Gaussian form $P(V) =
\frac{1}{\sqrt{2\pi s^2}}\text{exp}(-V^2/2 s^2)$ and the corresponding
density of states can be written as (also see Sec.~\ref{sec:bden}):
\begin{equation}
\begin{array}{l l l }
D_e(E)= \int_{-\infty}^{E} \dfrac{g_sg_v m}{2\pi \hbar^2} P(V)dV
\\
\\
= \dfrac{D_0}{2} \text{erfc}(-\dfrac{E}{\sqrt{2} s})
\end{array}
\end{equation}
where erfc$(x)$ is the complementary error function, $s = V_{rms}$ (as
given in Eq.~\ref{eq:flucV}), $D_0 =\dfrac{g_s g_v m}{2\pi \hbar^2}$,
$g_s=2$ and $g_v=2$ are the spin and valley degeneracies,
respectively. The main difference between MLG and BLG is
in their density of states of non-interacting systems. The
homogeneous  MLG system has the linear energy-dependent density of
states while the density of states of the homogeneous BLG is independent
of energy, which leads to different Thomas-Fermi screening
wavevectors. The potential fluctuation in Eq.~\ref{eq:flucV} affects the
electronic density of states in BLG. But the fluctuations depend on
the screening via Eq.~\ref{eq:Ak} while the screening depends on the
density of states via Eq.~\ref{eq:bs}. Therefore, we have a coupled
problem which must be solved self-consistently.


 In Fig.~\ref{fig:bparameter}, the broadening parameter $s$ and
 the screening constant $q_{TF}$ are plotted for various Fermi
 levels (i.e. the carrier density $n$). As shown for MLG, the BLG
parameters $(s, q_{TF})$ are also non-trivial function of
the fixed charge density $n_{imp}$,
 the SiO$_2$ thickness $d$, the location of the fixed charged
 impurity $z_0$, and  the Fermi level.
The different charged impurity configurations and
 carrier densities have similar effects on potential fluctuations of
 bilayer graphene as we discussed for monolayer graphene. The results
 for $s(q_{TF})$ are also quite similar to that of MLG (in
 Fig.~\ref{fig:parameter}) only with small numerical difference.

In Fig.~\ref{fig:bdosden}, the self-consistent electronic density of
states of BLG has been calculated using SiO$_2$ thickness $d=100$ nm
and distance of fixed charged impurities $z_0 = 1$ nm for different
impurity densities. The higher impurity density changes the density of
states more dramatically. In Fig.~\ref{fig:bdosE}, we show the
electronic density of states of BLG for different charged impurity
configurations and carrier densities. The existence of random charged
impurities reduces the electronic density of states for
$E>0$ but create a band tail for $E<0$.



\begin{thebibliography}{45}
\expandafter\ifx\csname natexlab\endcsname\relax\def\natexlab#1{#1}\fi
\expandafter\ifx\csname bibnamefont\endcsname\relax
  \def\bibnamefont#1{#1}\fi
\expandafter\ifx\csname bibfnamefont\endcsname\relax
  \def\bibfnamefont#1{#1}\fi
\expandafter\ifx\csname citenamefont\endcsname\relax
  \def\citenamefont#1{#1}\fi
\expandafter\ifx\csname url\endcsname\relax
  \def\url#1{\texttt{#1}}\fi
\expandafter\ifx\csname urlprefix\endcsname\relax\def\urlprefix{URL }\fi
\providecommand{\bibinfo}[2]{#2}
\providecommand{\eprint}[2][]{\url{#2}}

\bibitem[{\citenamefont{Das~Sarma et~al.}(2011)\citenamefont{Das~Sarma, Adam,
  Hwang, and Rossi}}]{dassarma2010}
\bibinfo{author}{\bibfnamefont{S.}~\bibnamefont{Das~Sarma}},
  \bibinfo{author}{\bibfnamefont{S.}~\bibnamefont{Adam}},
  \bibinfo{author}{\bibfnamefont{E.~H.} \bibnamefont{Hwang}}, \bibnamefont{and}
  \bibinfo{author}{\bibfnamefont{E.}~\bibnamefont{Rossi}},
  \bibinfo{journal}{Rev. Mod. Phys.} \textbf{\bibinfo{volume}{83}},
  \bibinfo{pages}{407} (\bibinfo{year}{2011}).

\bibitem[{\citenamefont{Castro et~al.}(2007)\citenamefont{Castro, Novoselov,
  Morozov, Peres, dos Santos, Nilsson, Guinea, Geim, and
  Castro~Neto}}]{castro2007}
\bibinfo{author}{\bibfnamefont{E.~V.} \bibnamefont{Castro}},
  \bibinfo{author}{\bibfnamefont{K.~S.} \bibnamefont{Novoselov}},
  \bibinfo{author}{\bibfnamefont{S.~V.} \bibnamefont{Morozov}},
  \bibinfo{author}{\bibfnamefont{N.~M.~R.} \bibnamefont{Peres}},
  \bibinfo{author}{\bibfnamefont{J.~M. B.~L.} \bibnamefont{dos Santos}},
  \bibinfo{author}{\bibfnamefont{J.}~\bibnamefont{Nilsson}},
  \bibinfo{author}{\bibfnamefont{F.}~\bibnamefont{Guinea}},
  \bibinfo{author}{\bibfnamefont{A.~K.} \bibnamefont{Geim}}, \bibnamefont{and}
  \bibinfo{author}{\bibfnamefont{A.~H.} \bibnamefont{Castro~Neto}},
  \bibinfo{journal}{Phys. Rev. Lett.} \textbf{\bibinfo{volume}{99}},
  \bibinfo{pages}{216802} (\bibinfo{year}{2007}).

\bibitem[{\citenamefont{Lee et~al.}(2010)\citenamefont{Lee, Bae, Jang, Jang,
  Zhu, Sim, Song, Hong, and Ahn}}]{Youngbin_NanoLet10}
\bibinfo{author}{\bibfnamefont{Y.}~\bibnamefont{Lee}},
  \bibinfo{author}{\bibfnamefont{S.}~\bibnamefont{Bae}},
  \bibinfo{author}{\bibfnamefont{H.}~\bibnamefont{Jang}},
  \bibinfo{author}{\bibfnamefont{S.}~\bibnamefont{Jang}},
  \bibinfo{author}{\bibfnamefont{S.-E.} \bibnamefont{Zhu}},
  \bibinfo{author}{\bibfnamefont{S.~H.} \bibnamefont{Sim}},
  \bibinfo{author}{\bibfnamefont{Y.~I.} \bibnamefont{Song}},
  \bibinfo{author}{\bibfnamefont{B.~H.} \bibnamefont{Hong}}, \bibnamefont{and}
  \bibinfo{author}{\bibfnamefont{J.-H.} \bibnamefont{Ahn}},
  \bibinfo{journal}{Nano Letters} \textbf{\bibinfo{volume}{10}},
  \bibinfo{pages}{490} (\bibinfo{year}{2010}).

\bibitem[{\citenamefont{Novoselov et~al.}(2005)\citenamefont{Novoselov, Jiang,
  Schedin, Booth, Khotkevich, Morozov, and Geim}}]{Novoselov}
\bibinfo{author}{\bibfnamefont{K.~S.} \bibnamefont{Novoselov}},
  \bibinfo{author}{\bibfnamefont{D.}~\bibnamefont{Jiang}},
  \bibinfo{author}{\bibfnamefont{F.}~\bibnamefont{Schedin}},
  \bibinfo{author}{\bibfnamefont{T.~J.} \bibnamefont{Booth}},
  \bibinfo{author}{\bibfnamefont{V.~V.} \bibnamefont{Khotkevich}},
  \bibinfo{author}{\bibfnamefont{S.~V.} \bibnamefont{Morozov}},
  \bibnamefont{and} \bibinfo{author}{\bibfnamefont{A.~K.} \bibnamefont{Geim}},
  \bibinfo{journal}{Proc.\ Natl.\ Acad.\ Sci.\ USA}
  \textbf{\bibinfo{volume}{102}}, \bibinfo{pages}{10451 (2005); K. S.
  Novoselov, A. K. Geim, S. V. Morozov, D. Jiang, M. I. Katsnelson, I. V.
  Grigorieva, S. V. Dubonos, and A. A. Firsov, Nature {\bf 438}, 197}
  (\bibinfo{year}{2005}).

\bibitem[{\citenamefont{Tan et~al.}(2007{\natexlab{a}})\citenamefont{Tan,
  Zhang, Bolotin, Zhao, Adam, Hwang, Das~Sarma, Stormer, and
  Kim}}]{TanDas_PRL07}
\bibinfo{author}{\bibfnamefont{Y.-W.} \bibnamefont{Tan}},
  \bibinfo{author}{\bibfnamefont{Y.}~\bibnamefont{Zhang}},
  \bibinfo{author}{\bibfnamefont{K.}~\bibnamefont{Bolotin}},
  \bibinfo{author}{\bibfnamefont{Y.}~\bibnamefont{Zhao}},
  \bibinfo{author}{\bibfnamefont{S.}~\bibnamefont{Adam}},
  \bibinfo{author}{\bibfnamefont{E.~H.} \bibnamefont{Hwang}},
  \bibinfo{author}{\bibfnamefont{S.}~\bibnamefont{Das~Sarma}},
  \bibinfo{author}{\bibfnamefont{H.~L.} \bibnamefont{Stormer}},
  \bibnamefont{and} \bibinfo{author}{\bibfnamefont{P.}~\bibnamefont{Kim}},
  \bibinfo{journal}{Phys. Rev. Lett.} \textbf{\bibinfo{volume}{99}},
  \bibinfo{pages}{246803} (\bibinfo{year}{2007}{\natexlab{a}}).

\bibitem[{\citenamefont{Chen et~al.}(2008{\natexlab{a}})\citenamefont{Chen,
  Jang, Adam, Fuhrer, Williams, and Ishigami}}]{Chen2008}
\bibinfo{author}{\bibfnamefont{J.~H.} \bibnamefont{Chen}},
  \bibinfo{author}{\bibfnamefont{C.}~\bibnamefont{Jang}},
  \bibinfo{author}{\bibfnamefont{S.}~\bibnamefont{Adam}},
  \bibinfo{author}{\bibfnamefont{M.}~\bibnamefont{Fuhrer}},
  \bibinfo{author}{\bibfnamefont{E.~D.} \bibnamefont{Williams}},
  \bibnamefont{and} \bibinfo{author}{\bibfnamefont{M.}~\bibnamefont{Ishigami}},
  \bibinfo{journal}{Nat. Phys.} \textbf{\bibinfo{volume}{4}},
  \bibinfo{pages}{377} (\bibinfo{year}{2008}{\natexlab{a}}).

\bibitem[{\citenamefont{Hong et~al.}(2009)\citenamefont{Hong, Zou, and
  Zhu}}]{ZhuExp_PRB09}
\bibinfo{author}{\bibfnamefont{X.}~\bibnamefont{Hong}},
  \bibinfo{author}{\bibfnamefont{K.}~\bibnamefont{Zou}}, \bibnamefont{and}
  \bibinfo{author}{\bibfnamefont{J.}~\bibnamefont{Zhu}},
  \bibinfo{journal}{Phys. Rev. B} \textbf{\bibinfo{volume}{80}},
  \bibinfo{pages}{241415} (\bibinfo{year}{2009}).

\bibitem[{\citenamefont{Chen et~al.}(2009)\citenamefont{Chen, Xia, and
  Tao}}]{Fang_Nano09}
\bibinfo{author}{\bibfnamefont{F.}~\bibnamefont{Chen}},
  \bibinfo{author}{\bibfnamefont{J.}~\bibnamefont{Xia}}, \bibnamefont{and}
  \bibinfo{author}{\bibfnamefont{N.}~\bibnamefont{Tao}}, \bibinfo{journal}{Nano
  Letters} \textbf{\bibinfo{volume}{9}}, \bibinfo{pages}{1621}
  (\bibinfo{year}{2009}).

\bibitem[{\citenamefont{Tan et~al.}(2007{\natexlab{b}})\citenamefont{Tan,
  Zhang, Stormer, and Kim}}]{YTan_EPJT07}
\bibinfo{author}{\bibfnamefont{Y.-W.} \bibnamefont{Tan}},
  \bibinfo{author}{\bibfnamefont{Y.}~\bibnamefont{Zhang}},
  \bibinfo{author}{\bibfnamefont{H.}~\bibnamefont{Stormer}}, \bibnamefont{and}
  \bibinfo{author}{\bibfnamefont{P.}~\bibnamefont{Kim}}, \bibinfo{journal}{Eur.
  Phys. J. Special Top.} \textbf{\bibinfo{volume}{148}}, \bibinfo{pages}{15}
  (\bibinfo{year}{2007}{\natexlab{b}}).

\bibitem[{\citenamefont{Chen et~al.}(2008{\natexlab{b}})\citenamefont{Chen,
  Jang, Xiao, Ishigami, and Fuhrer}}]{Chen_NPHNano2008}
\bibinfo{author}{\bibfnamefont{J.~H.} \bibnamefont{Chen}},
  \bibinfo{author}{\bibfnamefont{C.}~\bibnamefont{Jang}},
  \bibinfo{author}{\bibfnamefont{S.}~\bibnamefont{Xiao}},
  \bibinfo{author}{\bibfnamefont{M.}~\bibnamefont{Ishigami}}, \bibnamefont{and}
  \bibinfo{author}{\bibfnamefont{M.~S.} \bibnamefont{Fuhrer}},
  \bibinfo{journal}{Nat. Nanotechnol.} \textbf{\bibinfo{volume}{3}},
  \bibinfo{pages}{206} (\bibinfo{year}{2008}{\natexlab{b}}).

\bibitem[{\citenamefont{Hwang and Das~Sarma}(2009)}]{HwangScreen_PRB09}
\bibinfo{author}{\bibfnamefont{E.~H.} \bibnamefont{Hwang}} \bibnamefont{and}
  \bibinfo{author}{\bibfnamefont{S.}~\bibnamefont{Das~Sarma}},
  \bibinfo{journal}{Phys. Rev. B} \textbf{\bibinfo{volume}{79}},
  \bibinfo{pages}{165404} (\bibinfo{year}{2009}).

\bibitem[{\citenamefont{Lv and Wan}(2010)}]{Shaolong_PRB10}
\bibinfo{author}{\bibfnamefont{M.}~\bibnamefont{Lv}} \bibnamefont{and}
  \bibinfo{author}{\bibfnamefont{S.}~\bibnamefont{Wan}},
  \bibinfo{journal}{Phys. Rev. B} \textbf{\bibinfo{volume}{81}},
  \bibinfo{pages}{195409} (\bibinfo{year}{2010}).

\bibitem[{\citenamefont{Hwang et~al.}(2007)\citenamefont{Hwang, Adam, and
  Das~Sarma}}]{EHwang_PRL07}
\bibinfo{author}{\bibfnamefont{E.~H.} \bibnamefont{Hwang}},
  \bibinfo{author}{\bibfnamefont{S.}~\bibnamefont{Adam}}, \bibnamefont{and}
  \bibinfo{author}{\bibfnamefont{S.}~\bibnamefont{Das~Sarma}},
  \bibinfo{journal}{Phys. Rev. Lett.} \textbf{\bibinfo{volume}{98}},
  \bibinfo{pages}{186806} (\bibinfo{year}{2007}).

\bibitem[{\citenamefont{Rossi et~al.}(2009)\citenamefont{Rossi, Adam, and
  Das~Sarma}}]{RossiAdamDas_PRB09}
\bibinfo{author}{\bibfnamefont{E.}~\bibnamefont{Rossi}},
  \bibinfo{author}{\bibfnamefont{S.}~\bibnamefont{Adam}}, \bibnamefont{and}
  \bibinfo{author}{\bibfnamefont{S.}~\bibnamefont{Das~Sarma}},
  \bibinfo{journal}{Phys. Rev. B} \textbf{\bibinfo{volume}{79}},
  \bibinfo{pages}{245423} (\bibinfo{year}{2009}).

\bibitem[{\citenamefont{Martin et~al.}(2008)\citenamefont{Martin, Akerman,
  Ulbricht, T.Lohmann, Smet, Klitzing, and A.Yacoby}}]{Martin_NP08}
\bibinfo{author}{\bibfnamefont{J.}~\bibnamefont{Martin}},
  \bibinfo{author}{\bibfnamefont{N.}~\bibnamefont{Akerman}},
  \bibinfo{author}{\bibfnamefont{G.}~\bibnamefont{Ulbricht}},
  \bibinfo{author}{\bibnamefont{T.Lohmann}},
  \bibinfo{author}{\bibfnamefont{J.~H.} \bibnamefont{Smet}},
  \bibinfo{author}{\bibfnamefont{K.~V.} \bibnamefont{Klitzing}},
  \bibnamefont{and} \bibinfo{author}{\bibnamefont{A.Yacoby}},
  \bibinfo{journal}{Nat. Phys.} \textbf{\bibinfo{volume}{4}},
  \bibinfo{pages}{144} (\bibinfo{year}{2008}).

\bibitem[{\citenamefont{Zhang et~al.}(2011)\citenamefont{Zhang, Brar, Girit,
  Zettl, and Crommie}}]{YZhang_NP09}
\bibinfo{author}{\bibfnamefont{Y.}~\bibnamefont{Zhang}},
  \bibinfo{author}{\bibfnamefont{V.~W.} \bibnamefont{Brar}},
  \bibinfo{author}{\bibfnamefont{C.}~\bibnamefont{Girit}},
  \bibinfo{author}{\bibfnamefont{A.}~\bibnamefont{Zettl}}, \bibnamefont{and}
  \bibinfo{author}{\bibfnamefont{M.~F.} \bibnamefont{Crommie}},
  \bibinfo{journal}{Nat. Phys.} \textbf{\bibinfo{volume}{5}},
  \bibinfo{pages}{722 (2009); A. Deshpande, W. Bao, F. Miao, C. N. Lau, and B.
  J. LeRoy, Phys. Rev. B {\bf 79}, 205411 (2009); A. Deshpande, W. Bao, Z.
  Zhao, C. N. Lau, and B. J. LeRoy, Phys. Rev. B {\bf 83}, 155409}
  (\bibinfo{year}{2011}).

\bibitem[{\citenamefont{Heo et~al.}(2011)\citenamefont{Heo, Chung, Lee, Yang,
  Seo, Shin, Chung, Seo, Hwang, and Das~Sarma}}]{Heo_arXiv10}
\bibinfo{author}{\bibfnamefont{J.}~\bibnamefont{Heo}},
  \bibinfo{author}{\bibfnamefont{H.~J.} \bibnamefont{Chung}},
  \bibinfo{author}{\bibfnamefont{S.-H.} \bibnamefont{Lee}},
  \bibinfo{author}{\bibfnamefont{H.}~\bibnamefont{Yang}},
  \bibinfo{author}{\bibfnamefont{D.~H.} \bibnamefont{Seo}},
  \bibinfo{author}{\bibfnamefont{J.~K.} \bibnamefont{Shin}},
  \bibinfo{author}{\bibfnamefont{U.-I.} \bibnamefont{Chung}},
  \bibinfo{author}{\bibfnamefont{S.}~\bibnamefont{Seo}},
  \bibinfo{author}{\bibfnamefont{E.~H.} \bibnamefont{Hwang}}, \bibnamefont{and}
  \bibinfo{author}{\bibfnamefont{S.}~\bibnamefont{Das~Sarma}},
  \bibinfo{journal}{Phys. Rev. B} \textbf{\bibinfo{volume}{84}},
  \bibinfo{pages}{035421} (\bibinfo{year}{2011}).

\bibitem[{\citenamefont{Zhu et~al.}(2009)\citenamefont{Zhu, Perebeinos,
  Freitag, and Avouris}}]{zhu2009}
\bibinfo{author}{\bibfnamefont{W.}~\bibnamefont{Zhu}},
  \bibinfo{author}{\bibfnamefont{V.}~\bibnamefont{Perebeinos}},
  \bibinfo{author}{\bibfnamefont{M.}~\bibnamefont{Freitag}}, \bibnamefont{and}
  \bibinfo{author}{\bibfnamefont{P.}~\bibnamefont{Avouris}},
  \bibinfo{journal}{Phys. Rev. B} \textbf{\bibinfo{volume}{80}},
  \bibinfo{pages}{235402} (\bibinfo{year}{2009}).

\bibitem[{\citenamefont{Feldman et~al.}(2009)\citenamefont{Feldman, Martin, and
  Yacoby}}]{feldman2009}
\bibinfo{author}{\bibfnamefont{B.}~\bibnamefont{Feldman}},
  \bibinfo{author}{\bibfnamefont{J.}~\bibnamefont{Martin}}, \bibnamefont{and}
  \bibinfo{author}{\bibfnamefont{A.}~\bibnamefont{Yacoby}},
  \bibinfo{journal}{Nat. Phys.} \textbf{\bibinfo{volume}{5}},
  \bibinfo{pages}{889} (\bibinfo{year}{2009}).

\bibitem[{\citenamefont{Zou and Zhu}(2010)}]{zhujun_PRBR10}
\bibinfo{author}{\bibfnamefont{K.}~\bibnamefont{Zou}} \bibnamefont{and}
  \bibinfo{author}{\bibfnamefont{J.}~\bibnamefont{Zhu}},
  \bibinfo{journal}{Phys. Rev. B} \textbf{\bibinfo{volume}{82}},
  \bibinfo{pages}{081407} (\bibinfo{year}{2010}).

\bibitem[{\citenamefont{Nam et~al.}(2010)\citenamefont{Nam, Ki, and
  Lee}}]{Nam_PRB10}
\bibinfo{author}{\bibfnamefont{S.-G.} \bibnamefont{Nam}},
  \bibinfo{author}{\bibfnamefont{D.-K.} \bibnamefont{Ki}}, \bibnamefont{and}
  \bibinfo{author}{\bibfnamefont{H.-J.} \bibnamefont{Lee}},
  \bibinfo{journal}{Phys. Rev. B} \textbf{\bibinfo{volume}{82}},
  \bibinfo{pages}{245416} (\bibinfo{year}{2010}).

\bibitem[{\citenamefont{Hwang and
  Das~Sarma}(2008{\natexlab{a}})}]{HwangDasPhonon_PRB08}
\bibinfo{author}{\bibfnamefont{E.~H.} \bibnamefont{Hwang}} \bibnamefont{and}
  \bibinfo{author}{\bibfnamefont{S.}~\bibnamefont{Das~Sarma}},
  \bibinfo{journal}{Phys. Rev. B} \textbf{\bibinfo{volume}{77}},
  \bibinfo{pages}{115449} (\bibinfo{year}{2008}{\natexlab{a}}).

\bibitem[{\citenamefont{Min et~al.}(2011)\citenamefont{Min, Hwang, and
  Das~Sarma}}]{MinHwang_PRB11}
\bibinfo{author}{\bibfnamefont{H.}~\bibnamefont{Min}},
  \bibinfo{author}{\bibfnamefont{E.~H.} \bibnamefont{Hwang}}, \bibnamefont{and}
  \bibinfo{author}{\bibfnamefont{S.}~\bibnamefont{Das~Sarma}},
  \bibinfo{journal}{Phys. Rev. B} \textbf{\bibinfo{volume}{83}},
  \bibinfo{pages}{161404} (\bibinfo{year}{2011}).

\bibitem[{\citenamefont{Hwang and Das~Sarma}(2010)}]{Hwang_InsuPRB2010}
\bibinfo{author}{\bibfnamefont{E.~H.} \bibnamefont{Hwang}} \bibnamefont{and}
  \bibinfo{author}{\bibfnamefont{S.}~\bibnamefont{Das~Sarma}},
  \bibinfo{journal}{Phys. Rev. B} \textbf{\bibinfo{volume}{82}},
  \bibinfo{pages}{081409} (\bibinfo{year}{2010}).

\bibitem[{\citenamefont{Abergel et~al.}(2011)\citenamefont{Abergel, Hwang, and
  Das~Sarma}}]{David_PRB10}
\bibinfo{author}{\bibfnamefont{D.~S.~L.} \bibnamefont{Abergel}},
  \bibinfo{author}{\bibfnamefont{E.~H.} \bibnamefont{Hwang}}, \bibnamefont{and}
  \bibinfo{author}{\bibfnamefont{S.}~\bibnamefont{Das~Sarma}},
  \bibinfo{journal}{Phys. Rev. B} \textbf{\bibinfo{volume}{83}},
  \bibinfo{pages}{085429} (\bibinfo{year}{2011}).

\bibitem[{\citenamefont{Rossi and Das~Sarma}(2008)}]{Rossi_PRL08}
\bibinfo{author}{\bibfnamefont{E.}~\bibnamefont{Rossi}} \bibnamefont{and}
  \bibinfo{author}{\bibfnamefont{S.}~\bibnamefont{Das~Sarma}},
  \bibinfo{journal}{Phys. Rev. Lett.} \textbf{\bibinfo{volume}{101}},
  \bibinfo{pages}{166803} (\bibinfo{year}{2008}).

\bibitem[{\citenamefont{Arnold}(1974)}]{Arnold_APL74}
\bibinfo{author}{\bibfnamefont{E.}~\bibnamefont{Arnold}},
  \bibinfo{journal}{Appl.\ Phys.\ Lett.} \textbf{\bibinfo{volume}{25}},
  \bibinfo{pages}{705} (\bibinfo{year}{1974}).

\bibitem[{\citenamefont{San-Jose et~al.}(2007)\citenamefont{San-Jose, Prada,
  and Golubev}}]{Jose_PRB07}
\bibinfo{author}{\bibfnamefont{P.}~\bibnamefont{San-Jose}},
  \bibinfo{author}{\bibfnamefont{E.}~\bibnamefont{Prada}}, \bibnamefont{and}
  \bibinfo{author}{\bibfnamefont{D.~S.} \bibnamefont{Golubev}},
  \bibinfo{journal}{Phys. Rev. B} \textbf{\bibinfo{volume}{76}},
  \bibinfo{pages}{195445 (2007); J. H. Bardarson, J. Tworzyd\l{}o, P. W.
  Brouwer and C. W. J. Beenakker, Phys. Rev. Lett. {\bf 99}, 106801 (2007); E.
  Louis, J. A. Verg\'es, F. Guinea, and G. Chiappe, Phys. Rev. B {\bf 75},
  085440} (\bibinfo{year}{2007}).

\bibitem[{\citenamefont{Zallen and Scher}(1971)}]{zallen1971}
\bibinfo{author}{\bibfnamefont{R.}~\bibnamefont{Zallen}} \bibnamefont{and}
  \bibinfo{author}{\bibfnamefont{H.}~\bibnamefont{Scher}},
  \bibinfo{journal}{Phys. Rev. B} \textbf{\bibinfo{volume}{4}},
  \bibinfo{pages}{4471} (\bibinfo{year}{1971}).

\bibitem[{\citenamefont{Eggarter and Cohen}(1970)}]{eggarter1970}
\bibinfo{author}{\bibfnamefont{T.~P.} \bibnamefont{Eggarter}} \bibnamefont{and}
  \bibinfo{author}{\bibfnamefont{M.~H.} \bibnamefont{Cohen}},
  \bibinfo{journal}{Phys. Rev. Lett.} \textbf{\bibinfo{volume}{25}},
  \bibinfo{pages}{807} (\bibinfo{year}{1970}).

\bibitem[{\citenamefont{Du et~al.}(2008)\citenamefont{Du, Skachko, Barker, and
  Andrei}}]{Du_NaNano08}
\bibinfo{author}{\bibfnamefont{X.}~\bibnamefont{Du}},
  \bibinfo{author}{\bibfnamefont{I.}~\bibnamefont{Skachko}},
  \bibinfo{author}{\bibfnamefont{A.}~\bibnamefont{Barker}}, \bibnamefont{and}
  \bibinfo{author}{\bibfnamefont{E.~Y.} \bibnamefont{Andrei}},
  \bibinfo{journal}{Nature Nanotech.} \textbf{\bibinfo{volume}{3}},
  \bibinfo{pages}{491} (\bibinfo{year}{2008}).

\bibitem[{\citenamefont{Bolotin et~al.}(2008)\citenamefont{Bolotin, Sikes,
  Hone, Stormer, and Kim}}]{Bolotin_RPL08}
\bibinfo{author}{\bibfnamefont{K.~I.} \bibnamefont{Bolotin}},
  \bibinfo{author}{\bibfnamefont{K.~J.} \bibnamefont{Sikes}},
  \bibinfo{author}{\bibfnamefont{J.}~\bibnamefont{Hone}},
  \bibinfo{author}{\bibfnamefont{H.~L.} \bibnamefont{Stormer}},
  \bibnamefont{and} \bibinfo{author}{\bibfnamefont{P.}~\bibnamefont{Kim}},
  \bibinfo{journal}{Phys. Rev. Lett.} \textbf{\bibinfo{volume}{101}},
  \bibinfo{pages}{096802} (\bibinfo{year}{2008}).

\bibitem[{\citenamefont{M\"uller et~al.}(2009)\citenamefont{M\"uller,
  Br\"auninger, and Trauzettel}}]{Muller_PRL09}
\bibinfo{author}{\bibfnamefont{M.}~\bibnamefont{M\"uller}},
  \bibinfo{author}{\bibfnamefont{M.}~\bibnamefont{Br\"auninger}},
  \bibnamefont{and}
  \bibinfo{author}{\bibfnamefont{B.}~\bibnamefont{Trauzettel}},
  \bibinfo{journal}{Phys. Rev. Lett.} \textbf{\bibinfo{volume}{103}},
  \bibinfo{pages}{196801} (\bibinfo{year}{2009}).

\bibitem[{\citenamefont{Kirkpatrick}(1973)}]{kirkpatrick1973}
\bibinfo{author}{\bibfnamefont{S.}~\bibnamefont{Kirkpatrick}},
  \bibinfo{journal}{Rev. Mod. Phys.} \textbf{\bibinfo{volume}{45}},
  \bibinfo{pages}{574} (\bibinfo{year}{1973}).

\bibitem[{\citenamefont{Das~Sarma et~al.}(2010)\citenamefont{Das~Sarma, Hwang,
  and Rossi}}]{DasEnrico_PRB10}
\bibinfo{author}{\bibfnamefont{S.}~\bibnamefont{Das~Sarma}},
  \bibinfo{author}{\bibfnamefont{E.~H.} \bibnamefont{Hwang}}, \bibnamefont{and}
  \bibinfo{author}{\bibfnamefont{E.}~\bibnamefont{Rossi}},
  \bibinfo{journal}{Phys. Rev. B} \textbf{\bibinfo{volume}{81}},
  \bibinfo{pages}{161407} (\bibinfo{year}{2010}).

\bibitem[{\citenamefont{Hwang and Das~Sarma}(2007)}]{HwangDas_PRB07}
\bibinfo{author}{\bibfnamefont{E.~H.} \bibnamefont{Hwang}} \bibnamefont{and}
  \bibinfo{author}{\bibfnamefont{S.}~\bibnamefont{Das~Sarma}},
  \bibinfo{journal}{Phys. Rev. B} \textbf{\bibinfo{volume}{75}},
  \bibinfo{pages}{205418} (\bibinfo{year}{2007}).

\bibitem[{\citenamefont{Ferreira et~al.}(2011)\citenamefont{Ferreira,
  Viana-Gomes, Nilsson, Mucciolo, Peres, and Castro~Neto}}]{Ferreira_PRB11}
\bibinfo{author}{\bibfnamefont{A.}~\bibnamefont{Ferreira}},
  \bibinfo{author}{\bibfnamefont{J.}~\bibnamefont{Viana-Gomes}},
  \bibinfo{author}{\bibfnamefont{J.}~\bibnamefont{Nilsson}},
  \bibinfo{author}{\bibfnamefont{E.~R.} \bibnamefont{Mucciolo}},
  \bibinfo{author}{\bibfnamefont{N.~M.~R.} \bibnamefont{Peres}},
  \bibnamefont{and} \bibinfo{author}{\bibfnamefont{A.~H.}
  \bibnamefont{Castro~Neto}}, \bibinfo{journal}{Phys. Rev. B}
  \textbf{\bibinfo{volume}{83}}, \bibinfo{pages}{165402}
  (\bibinfo{year}{2011}).

\bibitem[{\citenamefont{Hwang and Das~Sarma}(2008{\natexlab{b}})}]{EH_PRL08}
\bibinfo{author}{\bibfnamefont{E.~H.} \bibnamefont{Hwang}} \bibnamefont{and}
  \bibinfo{author}{\bibfnamefont{S.}~\bibnamefont{Das~Sarma}},
  \bibinfo{journal}{Phys. Rev. Lett.} \textbf{\bibinfo{volume}{101}},
  \bibinfo{pages}{156802} (\bibinfo{year}{2008}{\natexlab{b}}).

\bibitem[{\citenamefont{Adam et~al.}(2007)\citenamefont{Adam, Hwang, Galitski,
  and Das~Sarma}}]{Adam_PNAS07}
\bibinfo{author}{\bibfnamefont{S.}~\bibnamefont{Adam}},
  \bibinfo{author}{\bibfnamefont{E.~H.} \bibnamefont{Hwang}},
  \bibinfo{author}{\bibfnamefont{V.~M.} \bibnamefont{Galitski}},
  \bibnamefont{and}
  \bibinfo{author}{\bibfnamefont{S.}~\bibnamefont{Das~Sarma}},
  \bibinfo{journal}{Proc.\ Natl.\ Acad.\ Sci.\ USA}
  \textbf{\bibinfo{volume}{104}}, \bibinfo{pages}{18392}
  (\bibinfo{year}{2007}).

\bibitem[{\citenamefont{Morozov et~al.}(2008)\citenamefont{Morozov, Novoselov,
  Katsnelson, Schedin, Elias, Jaszczak, and Geim}}]{Morozov_RPL08}
\bibinfo{author}{\bibfnamefont{S.~V.} \bibnamefont{Morozov}},
  \bibinfo{author}{\bibfnamefont{K.~S.} \bibnamefont{Novoselov}},
  \bibinfo{author}{\bibfnamefont{M.~I.} \bibnamefont{Katsnelson}},
  \bibinfo{author}{\bibfnamefont{F.}~\bibnamefont{Schedin}},
  \bibinfo{author}{\bibfnamefont{D.~C.} \bibnamefont{Elias}},
  \bibinfo{author}{\bibfnamefont{J.~A.} \bibnamefont{Jaszczak}},
  \bibnamefont{and} \bibinfo{author}{\bibfnamefont{A.~K.} \bibnamefont{Geim}},
  \bibinfo{journal}{Phys. Rev. Lett.} \textbf{\bibinfo{volume}{100}},
  \bibinfo{pages}{016602} (\bibinfo{year}{2008}).

\bibitem[{\citenamefont{Xiao et~al.}(2010)\citenamefont{Xiao, Chen, Adam,
  Williams, and Fuhrer}}]{Shudong_PRB10}
\bibinfo{author}{\bibfnamefont{S.}~\bibnamefont{Xiao}},
  \bibinfo{author}{\bibfnamefont{J.-H.} \bibnamefont{Chen}},
  \bibinfo{author}{\bibfnamefont{S.}~\bibnamefont{Adam}},
  \bibinfo{author}{\bibfnamefont{E.~D.} \bibnamefont{Williams}},
  \bibnamefont{and} \bibinfo{author}{\bibfnamefont{M.~S.}
  \bibnamefont{Fuhrer}}, \bibinfo{journal}{Phys. Rev. B}
  \textbf{\bibinfo{volume}{82}}, \bibinfo{pages}{041406}
  (\bibinfo{year}{2010}).

\bibitem[{\citenamefont{Oostinga et~al.}(2008)\citenamefont{Oostinga, Heersche,
  Liu, Morpurgo, and Vandersypen}}]{oostinga2008}
\bibinfo{author}{\bibfnamefont{J.~B.} \bibnamefont{Oostinga}},
  \bibinfo{author}{\bibfnamefont{H.~B.} \bibnamefont{Heersche}},
  \bibinfo{author}{\bibfnamefont{X.}~\bibnamefont{Liu}},
  \bibinfo{author}{\bibfnamefont{A.~F.} \bibnamefont{Morpurgo}},
  \bibnamefont{and} \bibinfo{author}{\bibfnamefont{L.~M.~K.}
  \bibnamefont{Vandersypen}}, \bibinfo{journal}{Nat. Mater.}
  \textbf{\bibinfo{volume}{7}}, \bibinfo{pages}{151} (\bibinfo{year}{2008}).

\bibitem[{\citenamefont{Mak et~al.}(2009)\citenamefont{Mak, Lui, Shan, and
  Heinz}}]{mak2009}
\bibinfo{author}{\bibfnamefont{K.~F.} \bibnamefont{Mak}},
  \bibinfo{author}{\bibfnamefont{C.~H.} \bibnamefont{Lui}},
  \bibinfo{author}{\bibfnamefont{J.}~\bibnamefont{Shan}}, \bibnamefont{and}
  \bibinfo{author}{\bibfnamefont{T.~F.} \bibnamefont{Heinz}},
  \bibinfo{journal}{Phys. Rev. Lett.} \textbf{\bibinfo{volume}{102}},
  \bibinfo{pages}{256405} (\bibinfo{year}{2009}).

\bibitem[{\citenamefont{Stern}(1976)}]{Stern_SS76}
\bibinfo{author}{\bibfnamefont{F.}~\bibnamefont{Stern}},
  \bibinfo{journal}{Surface Science} \textbf{\bibinfo{volume}{58}},
  \bibinfo{pages}{162 } (\bibinfo{year}{1976}).

\bibitem[{\citenamefont{Stern and Howard}(1967)}]{Stern_PR67}
\bibinfo{author}{\bibfnamefont{F.}~\bibnamefont{Stern}} \bibnamefont{and}
  \bibinfo{author}{\bibfnamefont{W.~E.} \bibnamefont{Howard}},
  \bibinfo{journal}{Phys. Rev.} \textbf{\bibinfo{volume}{163}},
  \bibinfo{pages}{816} (\bibinfo{year}{1967}).

\end{thebibliography}
\end{document}